\documentclass[10pt, conference, a4paper]{IEEEtran}

\IEEEoverridecommandlockouts

\usepackage{cite}
\usepackage{amsmath,amssymb,amsfonts}
\usepackage[ruled,lined,linesnumbered]{algorithm2e}
\usepackage{algorithmic}
\usepackage{algorithm2e} 
\usepackage{graphicx}
\usepackage{textcomp}
\usepackage{xcolor}
\usepackage{caption}
\usepackage{subcaption}
\usepackage{float}
\usepackage{pifont}
\usepackage{comment}

\def\BibTeX{{\rm B\kern-.05em{\sc i\kern-.025em b}\kern-.08em
    T\kern-.1667em\lower.7ex\hbox{E}\kern-.125emX}}
    
\newcommand{\cmark}{\ding{51}}
\newcommand{\xmark}{\ding{55}}

\newcommand{\name}{OROS}
\newcommand{\benchmark}{Oracle}

\begin{document}

\title{OROS: Online Operation and Orchestration of Collaborative Robots using 5G}

 \author{\IEEEauthorblockN{
 Arnau Romero\IEEEauthorrefmark{1},
 Carmen Delgado\IEEEauthorrefmark{1},
 Lanfranco~Zanzi\IEEEauthorrefmark{2},
 Xi Li\IEEEauthorrefmark{2},
 Xavier~Costa-P\'erez\IEEEauthorrefmark{1}\IEEEauthorrefmark{2}\IEEEauthorrefmark{3}}
 \IEEEauthorblockA{
 \IEEEauthorrefmark{1}i2CAT Foundation, Barcelona, Spain.
 Email:\{name.surname\}@i2cat.net,}
 \IEEEauthorrefmark{2}NEC Laboratories Europe, Heidelberg, Germany.
 Email:\{name.surname\}@neclab.eu,\\
 \IEEEauthorrefmark{3} ICREA, Barcelona, Spain.
 }

\maketitle

\begin{abstract}
The 5G mobile networks extend the capability for supporting collaborative robot operations in outdoor scenarios. However, the restricted battery life of robots still poses a major obstacle to their effective implementation and utilization in real scenarios. One of the most challenging situations is the execution of mission-critical tasks that require the use of various on-board sensors to perform simultaneous localization and mapping (SLAM) of unexplored environments. Given the time-sensitive nature of these tasks, completing them in the shortest possible time is of the highest importance.
In this paper, we analyze the benefits of 5G-enabled collaborative robots by enhancing the intelligence of the robot operation through joint orchestration of Robot Operating System (ROS) and 5G resources for energy-saving goals, addressing the problem from both offline and online manners. We propose \emph{\name{}}, a novel orchestration approach that minimizes mission-critical task completion times as well as overall energy consumption of 5G-connected robots by jointly optimizing robotic navigation and sensing together with infrastructure resources.  
We validate our 5G-enabled collaborative framework by means of Matlab/Simulink, ROS software and Gazebo simulator.
Our results show an improvement between 3.65\% and 11.98\% in exploration task by exploiting 5G orchestration features for battery savings when using 3 robots.

\end{abstract}

\begin{IEEEkeywords}
5G, Orchestration, Robotics, Optimization
\end{IEEEkeywords}

\section{Introduction}

Robots have been designed to interact with unknown environments and act on behalf of humans to minimize the risk of accidents or injuries. Thanks to their rapid deployment and relatively low cost, mobile ground robots as well as Unmanned Aerial Vehicles (UAVs) have recently emerged as alternatives to address emergency and mission-critical scenarios~\cite{SARDO}~\cite{Public_safety}.
Such use-cases drive the evolution of simple remote-controlled robots into moving platforms equipped with dedicated operating systems, advanced computing capabilities and multiple communication modules, to support autonomous navigation and robot control tasks, which can be also aided by Artificial Intelligence (AI) based solutions to perform more accurate decisions thanks to real-time multi-sensor data streams.

Simultaneous localization and mapping (SLAM) and object recognition are only some examples of robot applications that may be needed during rescue operations and deployment of first-aid support~\cite{Multi_Robot_SLAM}.
Most of the modern solutions require the setup of dedicated local networks, e.g., WiFi-based, which limits the applicability of robot technology to only indoor environments~\cite{Past_Present_SLAM}. However, in outdoor environments, the WiFi technology can hardly offer high reliability, availability, and low-latency connectivity requirements of  robotic applications.

In this context, the next generation of mobile networks (5G) is envisioned as a key-enabler to provide the outdoor wireless communication and enhance the autonomy of robotic applications~\cite{5GvsLTE}. The ubiquitous connectivity, high bandwidth and the low latency access provided by the modern mobile technologies, together with the possibility to seamlessly exploit multi-access edge computing platforms (MEC) deployed at the edge of the networks, and cloud computing to host processing and analytic services as virtual or container-based instances, enable unprecedent levels of flexibility in the deployment of robotic applications which may efficiently off-load computing tasks to the edge and alleviate their energy consumption~\cite{Mission_Critical_Edge}\cite{E2E_Reliability}. 
In the 5G context, an orchestrating entity acts as a mediator in charge of guaranteeing the most efficient use of the infrastructure resources, while pursuing the satisfaction of heterogeneous communication requirements. Similarly, an orchestration entity can be adopted in the robot domain to control the multitude of operational modules and on-board sensors of one or multiple robots, e.g., to optimize energy consumption and robots task operations. 
However, in the state-of-the-art both orchestration entities work independently, with little to no awareness of each other during the operational phases. On the one hand, the robot domain assumes ubiquitous and unlimited resource availability, both from the networking and cloud resource perspective, which may lead to unexpected performance drops when dealing with wireless and virtualized environments, especially in heterogeneous multi-robot scenarios where they share the same network resources. On the other hand, the 5G domain operates without knowledge of the actual resource requirements from the robot domain, mainly by over-provisioning the resources. This introduces inefficient networking resource usage and, perhaps more importantly, waste of battery on the robots, making it hard for such energy-constrained devices to fulfill their tasks. 

In our previous work~\cite{Delgado22}, we propose to integrate the orchestration logic from the network infrastructure and the robot domains. We addressed the problem from an \emph{offline} optimization perspective, enabling information exchange between the robots and the hosting infrastructure. In this way, not only energy-aware decisions performed in the robotic domain may be tuned according to given infrastructure conditions, but also infrastructure resources can be re-configured to meet robot communication and energy requirements in an efficient manner. 
Despite providing significant insights on the achievable energy savings, the offline approach requires accurate assumptions, e.g., a-priori knowledge on the position of the obstacles, which may be unfeasible in realistic scenarios.
To fill this gap, in this work we extend our previous framework towards an \emph{online} approach, allowing \name{} to operate only using information collected by the robots in real-time. To further validate our proposal, we also include a more accurate battery discharge model, run the overall robot software stack in ROS, and make use of the Gazebo simulator to reproduce realistic scenarios. To do so, we use two different machines, one running Matlab/Simulink with the OROS real-time orchestrator, and the other running as robot client with the Gazebo simulator.

The main contributions of this paper include:
\begin{itemize}
\item We design a framework for the deployment of robotic applications with the 5G infrastructure, defining the interfaces and the interactions between the proposed joint orchestration logic with the individual robot orchestrator and the 5G orchestrator.
\item We propose an optimization approach combining the orchestration of the 5G mobile infrastructure including mobile edge jointly with the energy-aware optimization of the on-board robot sensor applications, considering both offline and online solving strategies.
\item We evaluate our proposed approach in ideal conditions involving good wireless channel statistics and linear battery charge depletion, showcasing the theoretical achievable gains deriving from the joint orchestration of the 5G infrastructure and robot applications, both in terms of exploration time and resulting energy savings.
\item We finally extend our study to realistic conditions, adopting state-of-the art emulation tools and accounting for non-linear battery discharge rates in robots as well as variable 5G New Radio simulated communication link, identifying the main performance gaps.
\end{itemize}

The remainder of the paper is structured as follows.
Sec.~\ref{sec:RelatedWork} summarizes related works in the field.
Sec.~\ref{sec:framework} presents the main building blocks of our architecture and describes the interaction among the different modules.
Sec.~\ref{sec:problem} formulates our optimization problem, detailing our model assumptions and solving strategies.
Sec.~\ref{sec:perf_eval} validates the design principles of our solution by means of a comprehensive simulation campaign in ideal and realistic conditions, including a non-linear dynamic battery discharge rates and 5G wireless communication link.
Finally, Sec.~\ref{sec:conclusion} concludes this paper and discusses future works.

\section{Related Work}
\label{sec:RelatedWork}
Several works in the literature investigate the adoption of mobile networks to control robots in outdoor scenarios. 
In~\cite{Voigtlnder2017} the authors propose a framework for the offloading of time-critical and computational exhaustive operations onto a distributed node architecture, where the communication between the robot and the cloud server is done via 5G. 

Despite offloading computationally intensive task would help robots to save energy,  in realistic environments the limited energy availability provided by on-board batteries still represents a major limitation. Ideally, robots would require the largest batteries in order to extend their mobility range. At the same time, heavier batteries would impact on their energy consumption rate, thus introducing a design trade-off~\cite{Albonico2021}.

Swanborn \textit{et al.} identify robot navigation as the main energy consumer~\cite{Swanborn2020} at runtime. Additionally, they identify secondary sources of energy consumption, such as inefficient hardware, inefficient management algorithms, idle times, operational inefficiencies (e.g., poor quality software that leads to unnecessary stops and/or turns, as well as sharp acceleration and deceleration), processing energy, and finally, unnecessary communication and wasting of sensor data acquisition. We highlight that our solution will positively affect the last two drawbacks.

To optimize the exploration of unknown areas, several energy-aware management schemes have been proposed in the literature. \cite{Rappaport2016} proposes an approach for energy efficient path planning during autonomous mobile robot exploration. The idea is for a unique robot to efficiently explore the environment and periodically return to the starting point of the exploration for recharging its battery. The periodicity depends on an adaptive threshold that concurrently considers the movement of the robot, its power consumption, and the current state of the environment. The authors focus on minimizing the overall travel path in order to minimize the energy consumption. 
Since exploring an area poses hard limits in a single robot setting, in their follow-up work they extended the idea to teams of coordinated robots sharing a limited number of Charging Points (CPs) while exploring a structured, unknown environment with unknown obstacles. The objective in this case aims at exploring the area as fast as possible~\cite{Rappaport2017}. Since it is infeasible to pre-compute an optimized schedule due to a limited time horizon, an energy-aware planner is used for adaptive decision-making on when and where to recharge. However, their results are limited to numerical simulations.

\begin{table*}[t]\centering
\caption{Comparison of Related Works on Robot Exploration Strategies}
\begin{tabular}{c c c c c c}
\hline
\textbf{Work} & \textbf{MultiRobot}& \textbf{Evaluation}& \textbf{Battery behaviour}& \textbf{Charging}& \textbf{5G Channel Conditions }  \\
\hline
 Rappaport \textit{et al.} \cite{Rappaport2016} & \xmark  & Matlab & Linear & \cmark & Not considered  \\
 Rappaport \textit{et al.} \cite{Rappaport2017} & \cmark  & Numerical & Linear & \cmark & Not considered \\
Benkrid \textit{et al.} \cite{Benkrid2019} & \cmark  & Matlab \& Two mobile robots & Linear & \xmark &  Not considered\\
Raunholt \textit{et al.} \cite{raunholt_2021} & \xmark  & Mobile robot & Realistic non-linear & \cmark & Indoor and Ideal channel conditions\\
 Offline OROS~\cite{Delgado22} & \cmark  & Numerical & Linear & \cmark & Constant good channel conditions (simulated)\\
 This work & \cmark & Matlab \& Gazebo simulator & Realistic non-linear & \cmark & Variable channel quality\\
\hline    
\label{tab:sota}
\end{tabular}
\end{table*}

Similarly, Benkrid \textit{et al.} investigate the problem of multi-robot exploration in unknown environments. In their work, they propose a decentralized coordination approach to minimize the exploration time while considering the total motion energy saving of the mobile robots~\cite{Benkrid2019}. The exploration target is defined as a segment of the environment including the frontiers between the unknown and the explored areas. Each robot evaluates its relative rank, and compares with the other robots of the fleet, while considering the energy consumption to reach this exploration target. As a result, the robot is assigned to the segment for which it has the lowest rank. 
They evaluate their proposal through simulation experiments as well as ROS-enabled robots. However, inaccuracies during the robot localization and the map generation are not considered. Additionally, they only consider scenarios with unlimited energy availability or limited energy without the possibility of recharging. 
In~\cite{raunholt_2021} the authors consider a mobile edge cloud planner using TCP/IP and 5G emulation to provide a navigation plan for indoor rechargeable robots in industrial scenarios.
They consider both cloud-based and robot on-board path planning, comparing the two approaches in terms of communication and control loop delay, but limiting their analysis to indoor and industrial scenarios. Table~\ref{tab:sota} summarizes and highlights the main differences in robot exploration and path planning strategies and evaluated scenarios in the works above and our proposed approach.

Nevertheless, none of the above works have considered networking aspects in their optimization, nor adopt the NFV approach for the virtualization and cloud deployments of robotic applications at the edge/cloud infrastructure so as to offload some heavy computation tasks from the robots to the networks. Recent works such as~\cite{DEEP_Platform} and~\cite{5G-Dive}, started exploring the benefits of adopting 5G and cloud-native deployment in robotic applications exploiting the offloading of computational tasks to fog, edge or cloud systems to build a cloud-to-things continuum. Despite their promising results, the current research is still mainly focusing on virtualization aspects, and on the development of orchestration platforms to automate the deployment and allocation of both networking and computing resources over the 5G infrastructure.
However, such orchestration platforms only control the infrastructure resources, independent of the use of robotic applications, which neither consider the internal logic of robotic applications nor decide any actions for the robots. To the best of our knowledge, this is the first work to jointly consider robot application operations and infrastructure resource orchestration, pursuing enhanced robot autonomy, collaboration, and energy consumption optimization.

\section{Framework Overview}
\label{sec:framework}

\begin{figure}[t!]
      \centering
      \includegraphics[trim = 2cm 1.9cm 4.5cm 0cm , clip, width=\columnwidth ]{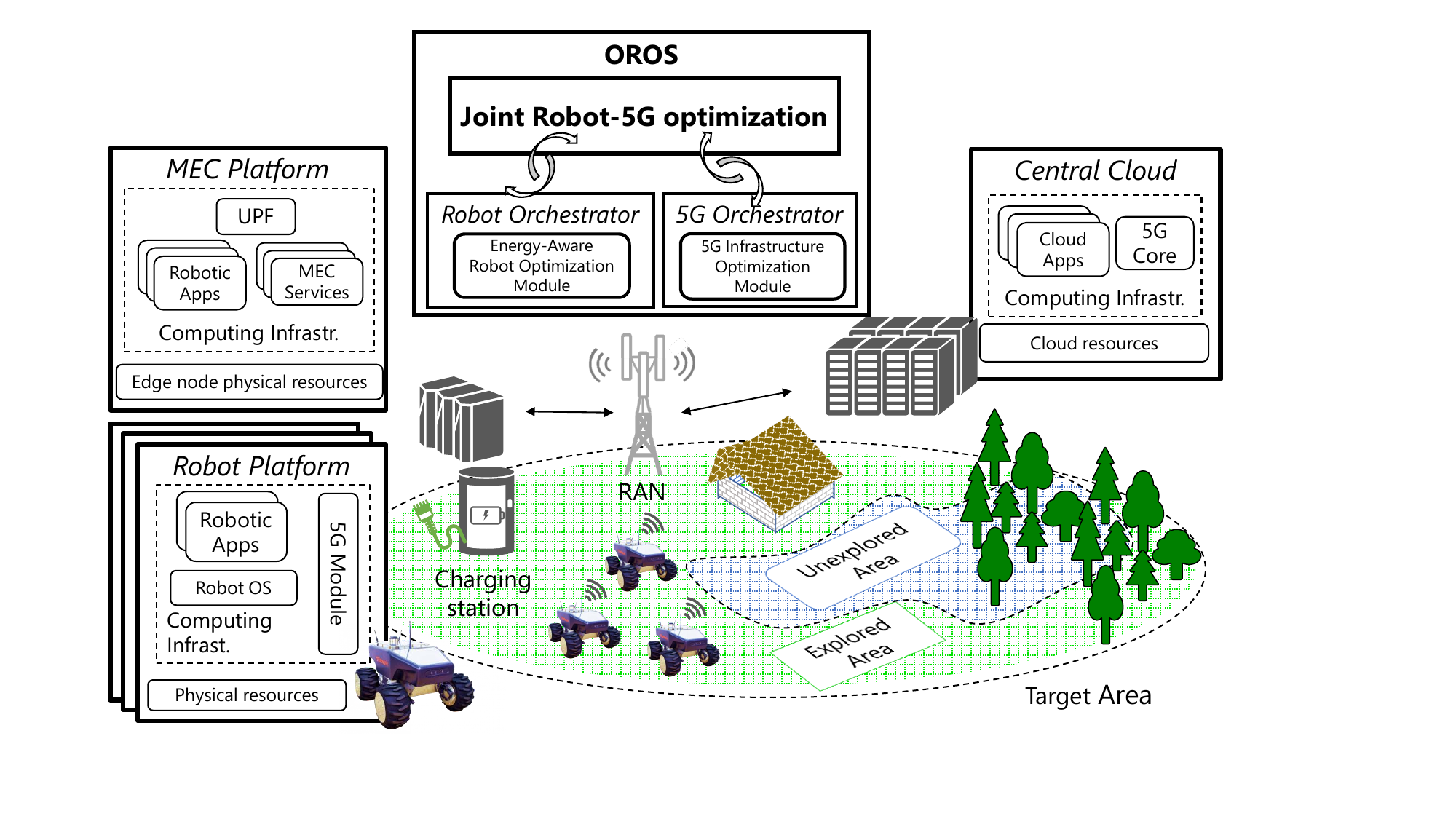}
      \caption{\small Overview of the architectural building blocks.}
      \label{fig:scenario}
      \vspace{-3mm}
\end{figure}

As depicted in Fig.~\ref{fig:scenario}, we consider a set of ground robots deployed in an unknown outdoor environment, where wireless communication is provided by means of a 5G network.
We assume a 5G Radio Access Network (RAN) composed by a set of base stations (gNBs) providing radio coverage over the area of interest. We also assume the presence of an edge and a remote cloud platforms, to host robotic applications and the 5G core functionalities, respectively, running as virtual or container-based instances within a computing infrastructure. A non-exhaustive list of robot applications include motion planning, video processing, etc., while the 5G core accounts for the set of user authentication, connection setup and mobility management functionalities proper of a mobile network. We assume the robot controller running in the edge premises, and rely on the presence of a User Plane Function (UPF) to seamlessly route the 5G data plane traffic generated by the robots to the edge platform, therefore favouring low-latency communication and wider bandwidth availability.

In this work, we focus on a surveillance (public protection and disaster relief) use-case. Cloud-based robots can perform 24/7 security inspections, replacing security personnel, reducing cost, and storing all data needed. They collect video and images and send them to the cloud for real-time identification of suspicious people and activity. Similar robots are already being used at airports and in outdoor rescue scenarios~\cite{airport_robots}. 

\subsection{Robot Orchestration}

In this paper, we build on the Robot Operating System (ROS) design and specifications~\cite{ROS} for the control and orchestration of robots and the onboarding of robotic applications. ROS is an open-source robotics middleware that provides common functionality (e.g., read sensor data, navigation, planning, etc.) over general hardware abstraction using low-level device control. It contains a collection of tools, libraries, APIs and conventions that simplify the task of creating complex and robust robot behavior across a variety of robotic systems~\cite{Quigley2009}. 
A system built using ROS consists of several processes, potentially running on a number of different hosts, connected at runtime in a peer-to-peer topology. The ROS topology is supported by a lookup mechanism to allow processes to find each other in real time. The mechanism is implemented through ROS Master and ROS nodes. 
While a ROS Master represents a stateless entity that coordinates the ROS nodes, the ROS nodes communicate with each other by passing messages, which are defined with a strictly typed data structure and published through topics. 
ROS was initially designed for standalone robotic applications, with the centralized design for serving a single robot in a local area network. 
As such, it is not suitable in upcoming multi-robot cloud-based applications, which demand for a distributed environment containing a variety of networking and cloud resources to be concurrently coordinated in order to connect and control multiple robots in potentially distributed areas. These problems have been partially addressed in the latest development efforts, which led to the ROS2 release. 
In its latest releases, ROS mitigates the issue of real-time topic sharing over a distributed platform by means of a novel Data Distributed Service (DDS)~\cite{Maruyama2016}, while supporting more flexible container-based deployments~\cite{ROS_Container}.
In order to seamlessly control robots and related robotic applications, we envision the \emph{Robot Orchestrator} entity as composed by a layered architecture consisting of three layers: application layer, ROS client layer, and ROS middleware layer. The application layer hosts a variety of robotic applications offered with run-time application programming capabilities. The ROS client layer provides a set of ROS client APIs~\cite{ROS_RCL_APIs} based on the built-in ROS client libraries to the developer interface supporting different languages such as C, C++, Python. The ROS middleware layer offers a set of APIs~\cite{ROS_RWP_APIs} to enable compatibility with different interchangeable low-level communication protocols, and support distributed data and service sharing. Through these provided APIs, the robot orchestrator is able to translate the application logic into a set of instructions to control and coordinate groups of robots via ROS command messages dispatched through its Southbound Interface (SBI).

\subsection{5G Orchestration}
In its simplest definition, a 5G Orchestrator is in charge of the allocation and management of the 5G infrastructure resources, including those required to enable robots communication and transmit their application data, as well as the set of computing resources to host and run the robotic control plane applications. For example, a 5G orchestration solution may decide on the amount of radio resources to be allocated in order to support both the robot control plane (e.g., navigation and velocity commands), and the robot data plane (e.g., video and sensor data) communications. Besides, the robot applications can be deployed in a virtualized environment, e.g., being containerized in a local or a cloud computing infrastructure. In such case, the 5G orchestrator is also in charge of the life-cycle management of such container-based robotic application instances, including their on-boarding, instantiation and termination, automatic scaling and self-healing operations. At last but not least, the 5G orchestrator can also determine a proper placement strategy of placing robotic applications which, thanks to softwarized and cloud-native approaches, may be deployed locally, i.e., onto the robot computing infrastructure, or remotely, in edge and cloud platforms, or even adopting hybrid approaches.
Nevertheless, an accurate placement strategy demands for proactive resource allocation in order to decide the optimal amount of computing, memory and storage resources to support the provisioning of various robotic applications, both onto the robots and edge/cloud platforms.
The 5G orchestrator can be built relying on existing open source orchestrator platforms such as Open Source MANO (OSM) or leveraging on open-source orchestration platforms developed for instance in~\cite{5GT_arch}~\cite{5Growth-commag}. From a system architecture perspective, the 5G orchestrator consists of three layers: service layer, orchestration layer and resource layer. The service layer defines an \textit{intent engine} to receive and process the application requests, translating the application requirements and mapping them to \emph{network slices} in the form of network slice template as defined by the 3GPP TS 28.531~\cite{ts28531} and the network slice resource model (3GPP TS 28.541~\cite{3GPPTS}). 
The orchestration layer consists of a \textit {Management and Network Orchestration (MANO)} stack to enforce the 5G policy on the allocation of resources, placement and life-cycle management of the robotic applications. These include the instantiation and releasing of computing resources to host drivers and processing instances managing the corresponding sensors on the robots, and their associated applications on a computing infrastructure when activating or deactivating sensors. On the bottom is the resource layer. It includes a \textit{Virtual Infrastructure Manager (VIM)}, which interacts with the underlying physical infrastructure and offers unified abstractions over the heterogeneous set of resources. It carries out monitoring, allocation and management of resources across the infrastructure and exposes this information to the orchestrator engine to guide its tasks. 

\subsection{\name{}}

\begin{figure}[t!]
      \centering
      \includegraphics[clip, trim = 5cm 0cm 8cm 1cm, width=\columnwidth]{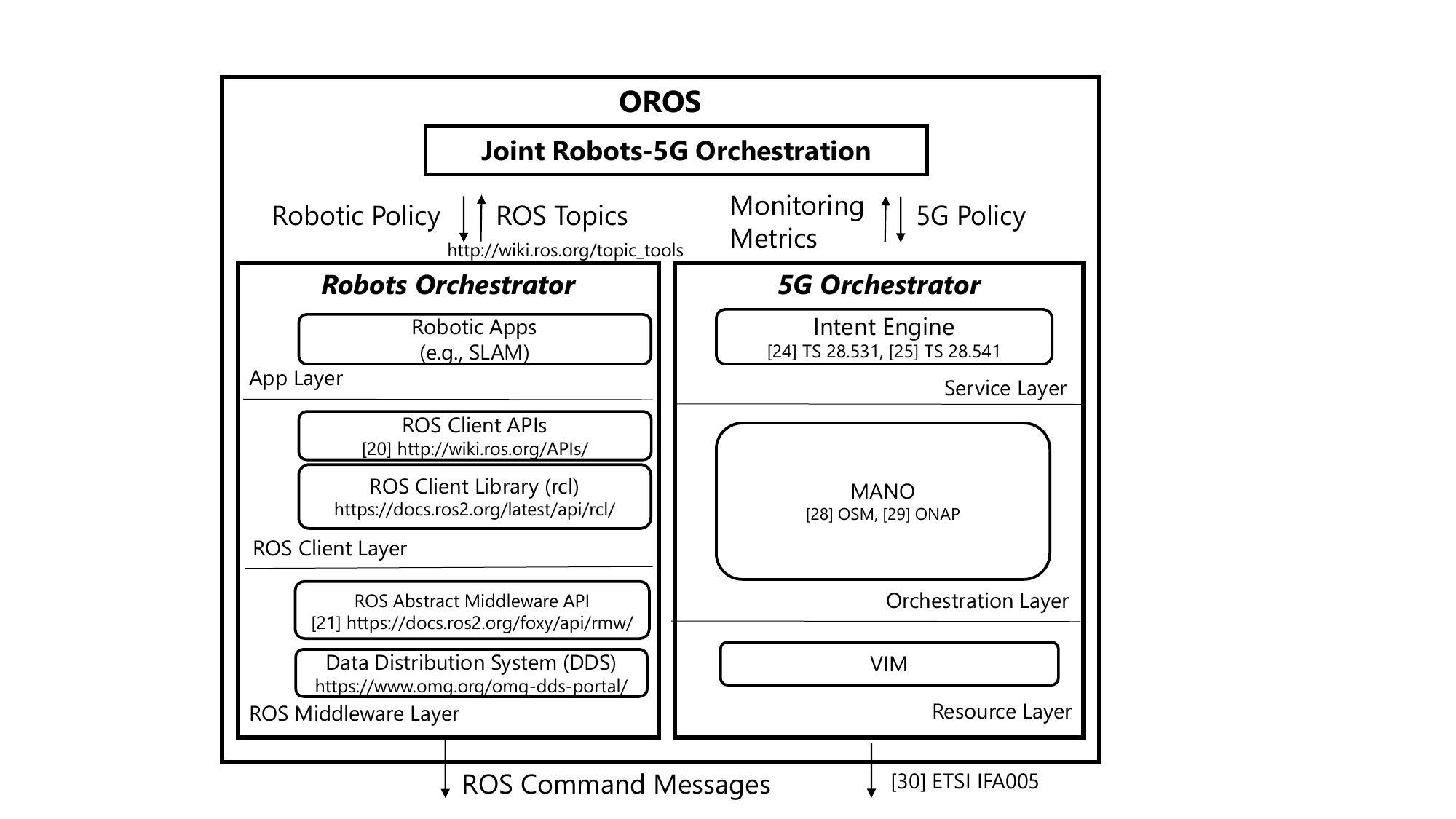}
      \caption{\small Architecture overview of the \name{} solution.}
      \vspace{-3mm}
      \label{fig:OROS}
\end{figure}

So far, major effort has been taken in the definition of orchestrating platforms in each corresponding domain, e.g., OSM~\cite{OSM}, ONAP~\cite{ONAP} in the 5G domain, and ROS~\cite{ROS} in the robotic environment. However, the orchestration tasks work independently, without taking consideration of the impact to each other on required resources to adapt to the robot requirements in real time, hence leading to inefficient use of resources and suboptimal robot actions. 
Traditionally, robot navigation and path planning processing are performed by a local robot controller deployed within the same premises.
When considering modern smart multi-robot platforms as well as novel use-cases brought by the Internet of Robotic Things (IoRT) paradigm ~\cite{IoT_robotics}, the ability to connect multiple robots, as well as stream, collect and analyze vast amounts of robotic data in real-time to powerful computing premises generally located in the edge/cloud, allows robots to offload demanding processing tasks to enable smarter robots that can autonomously adapt to changing conditions more quickly and accurately.
On the one hand, to enable the communication from the central controller entity to the individual robots, mobile network resources should be allocated and also adapted to guarantee the bandwidth and latency requirements, especially when computing offloading is also part of the robot's task.
On the other hand, enabled by the NFV technologies, modern robot sensing devices not only produce (and/or consume) sensor data, but also can be controlled by dedicated software applications (e.g., based on ROS) running on top of a shared computing infrastructure. Without a joint orchestration solution, these software instances will always remain active and consume the infrastructure resources, impacting on the overall energy consumption.
In this context, centralized edge/cloud computing platforms and mobile networks act as enablers in creating distributed robotic systems~\cite{cloud_architecture}. At the same time, decentralized architectures such as fog computing are also being considered, as to achieve a more scalable, efficient and effective robot resource management~\cite{fog_computing}.
Therefore, we advocate for the adoption of novel application-oriented orchestration solutions to guide the overall life-cycle management of cloud computing resources, provisioning of dedicated resources, and context-aware robot motion planning, pursuing energy savings strategies. 
To fill this gap, in this work we propose \name{}, a solution for the joint orchestration of the robotic and 5G ecosystem, to control ROS-driven collaborative connected robots in 5G networks.  The architecture design of the proposed \name{} solution is depicted in Fig.~\ref{fig:OROS}. \name{} seamlessly connects the orchestrating entities of the two domains, and coordinates their operations via a joint Robots-5G Orchestration module, as the central brain of our solution.

During the operation phase, assuming that the robotic applications are already instantiated and running, the joint Robots-5G orchestration module makes high-level joint orchestration decisions, namely \textit{Robotic Policies} and \textit{5G Policies}, and sends to each domain-specific orchestrator. It relies on the input taken from the monitoring metrics exposed by the Robot Orchestrator via its Northbound Interface (NBI) (e.g., ROS topics detailing velocity, location, and power consumption of the robots), and those exposed by the 5G Orchestrator (e.g., radio resource availability, and computing resources on the MEC and robot platforms for running the robotic applications, etc.).
Upon receiving the robotic policy, the Robot Orchestrator is in charge of reconfiguring the related robotic applications and translating the required corresponding actions to robot command messages, through exploiting the provided ROS client and middleware APIs. Examples of these include instructing the robot to move towards its new navigation goal, or switch off a specific sensor to save the energy.  In the 5G domain, once receiving a new 5G policy from the join optimization module, the Intent Engine will translate the received policy to update the corresponding network slices based on the new requirements. The 5G orchestration engine will further process the slice update request and optimize the reallocation of 5G resources (e.g., RAN, core, and MEC). This includes the instantiation and release of the networking and computing resources associated to the robotic applications, which can be (de)activated dynamically depending on the decision of \name{}. Moreover, the 5G orchestration decides on the placement and migration of the robotic applications towards the MEC and robot platforms. These decisions will be forwarded to the MANO to execute the operation workflows, and consequently to the VIM in order to enforce the configurations of the resources on the robot platforms and the 5G infrastructure via the SBI following ETSI IFA 005 specifications~\cite{ifa005}. In the following Sec.~\ref{sec:problem}, we detail the energy-aware mathematical formulation that guides the orchestration decisions while interacting with the specific domain orchestrators.

\begin{table*}[t]
\caption{Model parameters}
\label{tab:optParam}
\centering
\begin{tabular}{l|l}
\hline
\textbf{Parameter}	& \textbf{Definition}	\\
\hline
$\mathcal{T} = \{t_1, \dots, t_{|\mathcal{T}|} \}$		& Set of time instants; index $t$ refers to time instant $t_t$\\
$\mathcal{R}=\{r_1, \dots, r_{|\mathcal{R}|}\}$		& Set of robots; index $r$ refers to task $r_r$\\
$A \times B$                        & Geometric dimensions of the area of interest\\
$\mathcal{G} =\{g_{a,b}, \forall (a,b) \in (A,B) \}$ & Grid representing the area to be explored \\
$m_{a,b,a',b'}$ & Terrain-velocity constant for moving from position $g_{a,b}$ to $g_{a',b'}$  \\
$u_{r,t}$		& Binary decision variable indicating if the charging station is being used at time $t$ by robot $r$\\
$l_{r,t,a,b}$		& Binary decision variable indicating if robot $r$ is in position $g_{a,b}$ at time instant $t$\\
$e_{t,a,b}$		& Binary variable indicating if the unit of area $g_{a,b}$ has been explored at time $t$\\
$b_{r,t}$	    & Continuous variable indicating the battery level of robot $r$ at time instant $t$, where $0 \leq b_{r,t} \leq B_{max}$ \\
$CR$	        & Charging rate provided by the charging station \\
$P_{move_{a,b,a',b'}}$	    & Power consumed by moving from position $g_{a,b}$ to position $g_{a',b'}$\\
$P_{RX}$	    & Power consumed for receiving \\ 
$P_{TX,a,b}$	& Power consumed for transmitting \\ 
$P_{SEN}$	    & Power consumed by activating sensors, local data processing and on-robot computing infrastructure\\
\hline
\end{tabular}
\vspace{-5mm}
\end{table*}

\section{Problem Formulation}
\label{sec:problem}

\subsection{Energy-Aware Robot Orchestration Optimization}
\label{sec:Optimization}

Hereafter, we present our assumptions, notation and problem formulation to model the Energy-Aware Optimization problem. All variables and system parameters are resumed in Table~\ref{tab:optParam} to allow faster referencing.

\textbf{{Input variables}} Let us consider a discrete set of time instants denoted by the set $\mathcal{T} = \{t_1, \dots, t_{|\mathcal{T}|} \}$, and a set of robot devices $\mathcal{R}=\{r_1, \dots, r_{|\mathcal{R}|} \}$. Each robot $r \in \mathcal{R}$ is equipped with a rechargeable battery characterized by a limited capacity $B_{max}$,  $\forall r \in \mathcal{R}$,
whose charging status $b_{r,t}$ varies over time depending on the robot operational activities, i.e.,
$0 \leq b_{r,t} \leq B_{max} \quad  \forall t \in \mathcal{T}$. Each robot is also equipped with multiple sensors, such as cameras, Light Detection and Ranging (LiDAR) sensors, etc.

We assume robots to be deployed in an outdoor environment covered by a mobile infrastructure, as to enable 5G connectivity.
Without loss of generality, let us define the area of interest with dimensions $A \times B$ meters, and decompose the 2D surface into a grid $\mathcal{G} =\{g_{a,b}, \forall (a,b) \in (A,B) \}$,  where each element $g_{a,b} \in \mathcal{G}$ needs to be explored by at least one robot during the operational phase. The dimension of $g_{a,b}$, i.e., $|g_{a,b}|$, depends on the maximum field of view of the adopted robot camera and sensors, and we assume such cameras/sensors to be able to provide a $360^{\circ}$ view of the surrounding environment.

Each robot needs to receive periodical updates from the radio interface, as well as upload 
sensed environmental information.
For doing so, each robot adapts the Modulation and Coding Schemes (MCS) used according to the perceived channel quality, which we assume proportional to the instantaneous distance of the robot to the serving base station, as detailed in ~\cite{Delgado22}. We assume the serving base station located at position $(g_{a_{BS},b_{BS}}) \in \mathcal{G}$.

To keep track of the multi-robot exploration, we introduce $e_{t,a,b}$ as a binary variable indicating if the area unit $g_{a,b}$ has been already explored at time $t \in \mathcal{T}$, or not. Notably, the updates on the discovered area in previous time instants are inputs for the problem. Such information is shared between the robots and \name{} through update messages. 
Exploration of unknown areas may require charging stations to extend the operational time of rescue robots. Therefore, we assume a charging station (CS) located in position $(g_{a_{CS},b_{CS}}) \in \mathcal{G}$, which is known a-priori to all robots, which can be either available or unavailable (due to an already charging robot) at a certain time instant. The CS provides a charging rate $CR$ to recover the robot battery.
\begin{figure}[t!]
      \centering
      \includegraphics[trim = 0cm 0.5cm 0cm 0cm , clip, width=0.99\columnwidth ]{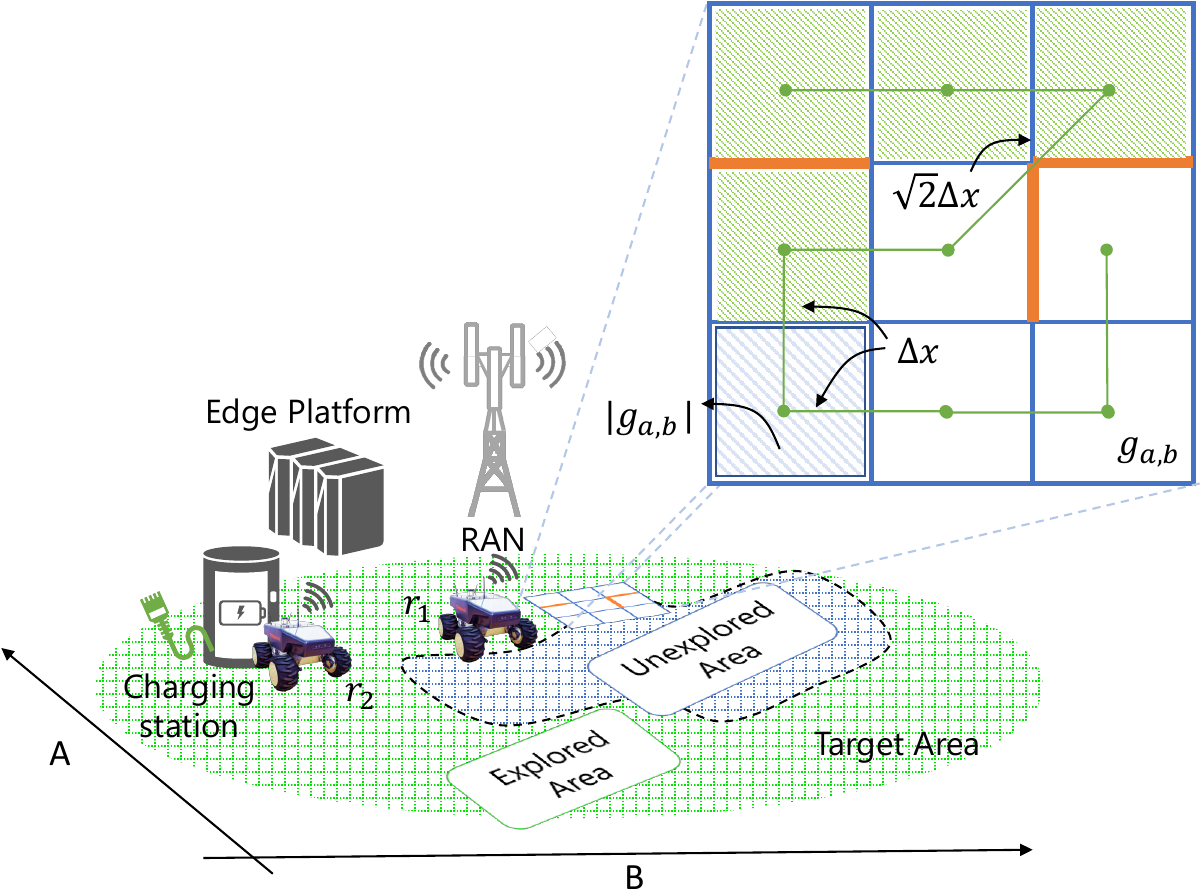}
      \caption{\small Example of robot exploration task solved with multiple robots.}
      \label{fig:example}
      \vspace{-3mm}
\end{figure}

We define $\Delta t$ as the period between two time instants in $\mathcal{T}$ (i.e., $\Delta t = |t_t - t_{t-1}|$), and assume the spatial mobility of every robot $r \in \mathcal{R}$ limited to a unit of distance $\Delta x$ (or $\sqrt{2}\Delta x$ if moving in diagonal) for every $\Delta t$, where $\Delta x$ is the distance between the centers of two neighbor area units $g_{a,b}$ and $g_{a',b'}$ as depicted in Fig.~\ref{fig:example}.
Therefore, it turns that each robot moves at a velocity \( \frac{\Delta x}{\Delta t} \) if moving into the left/right or up/down directions, but moves at velocity \( \frac{\sqrt{2}\Delta x }{\Delta t} \) when traveling in diagonal directions. Clearly, if the velocity increases, so does the corresponding energy cost. 
We consider this by introducing the terrain-velocity constant $m_{a,b,a',b'} \quad \forall a, a' \in A, \quad \forall b, b' \in B$, which weights the energy consumption of the robots from moving from position $g_{a,b}$ to position $g_{a',b'}$ \cite{Delgado22}. The value of $m_{a,b,a',b'}$ is dynamically set to infinite in case any of the robot detects an obstacle, therefore influencing the mobility of other robots in the corresponding unit of area in subsequent time intervals.
Whenever the robot moves, we can derive its power consumption by adapting Equation~1 of Rappaport~\cite{Rappaport2016}, where instead of only computing the velocity, we consider the terrain-velocity constant to account for obstacles and terrain conformity as:
\begin{equation}
	P_{move_{a,b,a',b'}} = 0.29 + 7.4 m_{a,b,a',b'}.
 	\label{eq:Pmove}
\end{equation}
As mentioned before, robots exploit an existing mobile infrastructure to communicate with the orchestrating entity. We also consider $P_{TXa,b}$ as a variable representing the energy consumed by the robot for transmitting data. The actual robot transmit power should be adapted as to compensate the radio path-loss. Therefore, the value of $P_{TXa,b}$ 
depends on the current robot location $(g_{a,b})$ and its distance from the serving base station, as later detailed in Sec.~\ref{sec:scenarioSetup}. 
Similarly, robots consume energy for receiving data. In this case, we consider $P_{RX}$ as a constant value independent of the robot location. 
Without loss of generality, we assume all packets to be of the same size.
Finally, since robots need to map and explore the terrain, they need to use their camera and sensors, as well as process those data to avoid inefficient raw data transmission. We collect the energy consumption derived by all these activities in the variable $P_{SEN}$, which represents the energy consumed by robot sensors and corresponding local data processing executed on the on-robot computing infrastructure.

\textbf{Decision variables}
Let $l_{r,t,a,b}$ be a binary decision variable to control the robot mobility. Its value gets positive if the robot $r$ is at position $g_{a,b}$ at time instant $t$.\\
Similarly, we introduce $u_{r,t}$ as a binary decision variable performing decisions on which robot $r$ has to recharge its battery at time $t$.

\textbf{Objective}
To explore an unknown area as fast as possible, and subsequently increase the chances of detecting the target object (or person), we need to maximize the explored area within the given time period $|\mathcal{T}| \times \Delta t$. 
Furthermore, we also want to ensure that the energy is consumed in the most efficient way, and for that reason, we include in the objective function the remaining battery of the robots in the last time instant. Therefore, we can write our objective function as:
\begin{equation}
	\label{eq:maximiza}
	\max \sum_{t \in \mathcal{T}} \sum_{(a,b) \in (A,B)} e_{t,a,b}  + \sigma \sum_{r \in \mathcal{R}}  b_{r,|\mathcal{T}|}
\end{equation}
where $\sigma$ is a scaling parameter that ensures the second term to be of comparable magnitude with the first term, while pursuing area exploration maximization.

\textbf{Constraints}
We assume that the charging station can only recharge one robot $r \in \mathcal{R}$ at every time instant $t \in \mathcal{T}$, therefore we introduce the following constraint:
\begin{equation}
	\label{eq:const1}
	\sum_{r \in \mathcal{R}} u_{r,t} \leq 1  \quad \forall t \in \mathcal{T},
\end{equation}
\noindent additionally, a robot $r \in \mathcal{R}$ can be charged at some time $t \in \mathcal{T}$ only if the robot was already in the charging station at $t-1$, and continues to stay there at time $t$:
\begin{equation}
	\label{eq:const6}
	u_{r,t} \leq  \frac{l_{r,t,a_{CS},b_{CS}} + l_{r,t-1,a_{CS},b_{CS}}}{2}   \quad \forall t \in \mathcal{T}, \forall r \in \mathcal{R}.
\end{equation}
\noindent Clearly, the duration of the recharge period may comprise multiple time intervals $t$, to allow more energy to fill the battery. With the following constraint, we ensure that each robot $r \in \mathcal{R}$ can only be in one place in every time instant  $t \in \mathcal{T}$:
\begin{equation}
	\label{eq:const7}
	\sum_{(a,b) \in (A,B)} l_{r,t,a,b} = 1  \quad \forall r \in \mathcal{R}, \forall t \in \mathcal{T}.
\end{equation}
\noindent In order to keep track of the exploration progress among multiple robots, if any robot $r \in \mathcal{R}$ visited an area unit $g_{a,b} \in (A, B)$ at some earlier time, or if it is exploring such area unit at the current time $t$, that area becomes explored at time $t$ and we update the variable $e_{t,a,b}$ accordingly.
\begin{equation}
	\label{eq:const3}
	e_{t,a,b} \leq 	e_{t-1,a,b} + \sum_{r \in \mathcal{R}}  l_{r,t,a,b}  \quad \forall t \in \mathcal{T}, \forall (a,b) \in (A,B),
\end{equation}
\begin{equation}
	\label{eq:const4}
	e_{t,a,b} \geq 		e_{t-1,a,b}  \quad \forall t \in \mathcal{T},    \forall (a,b) \in (A,B),  
\end{equation}
\begin{equation}
	\label{eq:const5}
	|\mathcal{R}| e_{t,a,b} \geq 		\sum_{r \in \mathcal{R}}  l_{r,t,a,b}   \quad \forall t \in \mathcal{T},    \forall (a,b) \in (A,B).  
\end{equation}
\noindent With the following constraint we define the mobility boundaries of the robots, and ensure that for each $\Delta t$ a robot $r \in \mathcal{R}$ can only move to a neighbor area unit, or stay in the same position.
For every time in $\mathcal{T}$ and algorithm execution, we consider the current robot locations as starting positions, while for the rest of the time instants, we have:
\begin{gather}
	l_{r,t+1,a,b} \leq l_{r,t,a,b} + l_{r,t,a-1,b} + l_{r,t,a+1,b} + l_{r,t,a,b-1} + \notag \\ 
	  l_{r,t,a,b+1} + 	l_{r,t,a-1,b-1} + l_{r,t,a+1,b+1} + l_{r,t,a-1,b+1} + \notag \\ 
 l_{r,t,a+1,b-1}  \quad \forall r \in \mathcal{R} , \forall t \in \mathcal{T}, \forall (a,b) \in (A,B).   \label{eq:const8} 
\end{gather}
\noindent As detailed in \cite{Delgado22}, Equation~\ref{eq:const8} affects the robot velocity. In fact, as mentioned before, the velocity and therefore the power consumption depends on the direction of the robot movement. This is taken into account in the constant $P_{move_{a,b,a',b'}}$.
Conversely, if the robot is recharging at the charging station (i.e., $u_{r,t} = 1$), the energy level of its battery increases with a constant charging rate $CR$. Furthermore, in every time instant the robot is not charging, it sends or receives data. When transmitting, the consumed power depends on the distance to the base station (according to $P_{TX,a,b}$). Robots report feedback information to the orchestrator, including current location, battery level, and detected obstacles. On the other hand, the consumed power of reception depends on $P_{RX}$ as previously discussed. Finally, if a robot has never been in an area unit, its sensors, camera, processing units and transmission elements should be active. However, if the robot is in an already explored area, an advanced orchestration solution may turn them off for the purpose of saving energy. Robots are expected to be reachable by the managing entity in every time instant. For this reason, we assume robots never switch off their receiving antennas.
Taking this into account, our algorithm updates the expected battery level $b_{r,t+1}$ by means of the following equation:

\begin{gather}
	b_{r,t+1} = b_{r,t} +  CR \times  u_{r,t+1}  - P_{RX} \times (1 - u_{r,t+1}) - \notag \\ 
	 \sum_{(a,b) \in (A,B)}  	\sum_{(a',b') \in (A,B)} l_{r,t,a,b} \times l_{r,t+1,a',b'} \times  P_{move_{a,b,a',b'}}  \notag \\ 
	- P_{SEN}  \times  \sum_{(a,b) \in (A,B)} (1 - e_{t,a,b}) \times l_{r,t+1,a,b} -  \label{eq:const9a}\\  
	\sum_{(a,b) \in (A,B)}   P_{TX,a,b} \times (1 - e_{t,a,b}) \times l_{r,t+1,a,b} \quad \forall t \in \mathcal{T} , \forall r \in \mathcal{R}. \notag
\end{gather}

\textbf{Outputs} An optimal solution of the problem would consider multiple outputs. First, it derives the optimal position of the robots, i.e, their path planning. Second, it specifies in which time instant each robot should charge. Third, it takes decisions on the state of the sensors, indicating whether they should be turned on or off at every specific time instant.
Finally, and as a way to compensate for non-flat exploration areas which may cause larger energy consumption, it provides an estimation of the battery charge level of each robot in the following instants of time. For example, if the estimated charge is considerably higher than the one reported by the robot, it could mean that the robot has gone through an uphill path.
The overall problem formulation can be summarized as follows: 

\noindent \textbf{Problem}~\texttt{OROS ($ \mathcal{T}$)} :
\label{prob:Oracle}
\begin{flalign}
  \quad\quad & \text{max} 	\sum_{t \in \mathcal{T}} \sum_{(a,b) \in (A,B)} e_{t,a,b}  + \sigma \sum_{r \in \mathcal{R}}  b_{r,|\mathcal{T}|} \nonumber & &\\
  \quad\quad & \text{subject to:} \nonumber\\
  \quad\quad & \quad\quad (\ref{eq:const1}) (\ref{eq:const6}) (\ref{eq:const7}) (\ref{eq:const3}) (\ref{eq:const4}) (\ref{eq:const5})  (\ref{eq:const8}) (\ref{eq:const9a}); \nonumber
\end{flalign}

\subsection{Offline vs Online Optimization}
\label{sec:offline_vs_online}

\begin{figure}[t!]
      \centering
      \includegraphics[trim = 0cm 0cm 0cm 0cm , clip, width=0.8\columnwidth ]{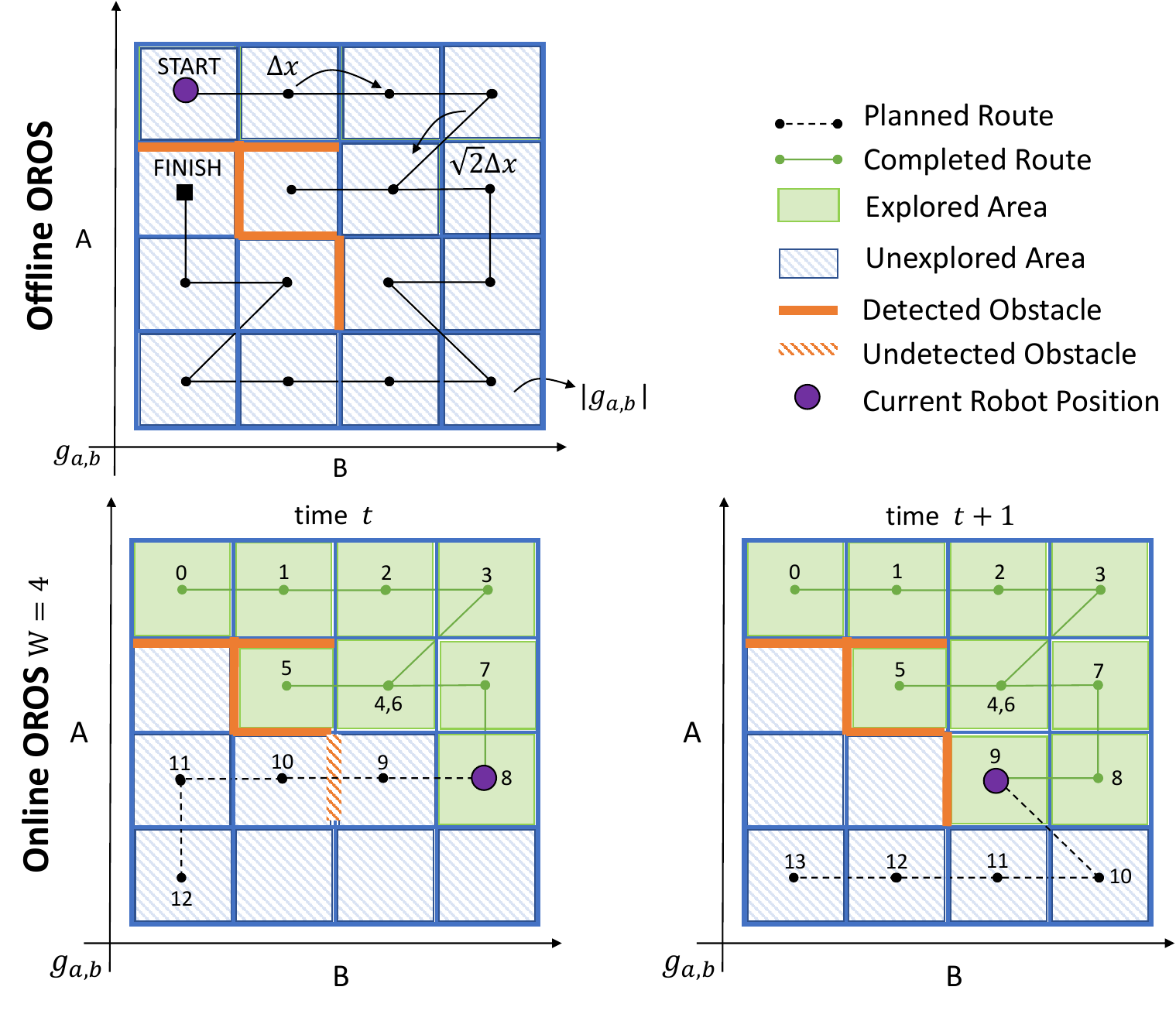}
\caption{\small Example of robot exploration problem solved with the state-of-the-art Offline (top) vs. proposed Online approach (bottom).}
      \label{fig:OfflineVsOnline}
\end{figure}

The problem described in Sec.~\ref{sec:Optimization} can be solved in different ways. For example, an \emph{offline} approach would have to consider complete knowledge over the input variables, and solve the problem over the whole time period as depicted in Fig.~\ref{fig:OfflineVsOnline} (upper row). 
Conversely, an \emph{online} algorithm would solve an instantaneous instance of the problem over multiple steps, one for every decision interval $t$, relying only on the limited amount of information dynamically collected by the robots, as depicted in Fig.~\ref{fig:OfflineVsOnline} (lower row).
Notably, in general, the output of the online algorithm may differ from the one of the corresponding \emph{offline} version, and not necessarily reach the same solution in presence of multiple local optimal points. 
In our previous work~\cite{Delgado22}, we solve this optimization task in an \emph{offline} manner, assuming a-priori and complete knowledge on the obstacle position and about the geographical landscape. We refer to this approach as \emph{Offline Oracle}. Algorithm \ref{alg:offline} summarizes its implementation as pseudocode. Although this approach represents a good benchmark for the best achievable performances, its application to real world scenarios is limited to very specific situations where input information regarding obstacle location can be known in advance and made available at startup.
Conversely, in this work, we develop an \emph{online} algorithm that considers only limited and up-to-dated information about the environment, i.e., those that are collected as feedbacks from the robots.
Every robot-related event, e.g., object detection, terrain slopes, etc. triggers a new optimization task as to devise the best navigation path and sensor management strategy based on the latest collected information. 
Despite this approach may intuitively provide enhanced performances, it may actually lead to unnecessary computation steps when considering only short-term planning.
Therefore, we propose a version of the solver that takes decisions in $\mathcal{W} = \{ t, \dots, t+W \} \subseteq \mathcal{T}$, i.e., the subset of the following $W$ time steps, being $W$ a configurable parameter describing the size of the decision time window.
Such approach is well-suited for real-time operations and allows for more efficient usage of computing resources.
Algorithm \ref{alg:online} shows how this problem should be executed. Notably, while Algorithm~\ref{alg:offline}  \texttt{Offline Oracle} is only executed once and assumes the variable $m_{a,b,a',b'}$ as a constant input to the problem, Algorithm~\ref{alg:online} \texttt{Online OROS} runs multiple times depending on both robot detection events in the current location $(a,b)$ and the value of $W$. The value of $m_{a,b,a',b'}$ gets updated at every iteration depending on the feedbacks from the robot. This affects the expected robot energy consumption due to mobility $P_{move_{a,b,a',b'}}$, as well as influences the transmission power $P_{TX,a,b}$ as later detailed in Sec.~\ref{sec:scenarioSetup}.
In the following, we evaluate the performances of these different approaches under a testing scenario through experiments over a simulated setup.

\begin{algorithm}[!t]
\small
\SetKwInOut{Input}{Input}
\SetKwInOut{Output}{Output}
\SetKwInOut{Return}{return}
\SetKwInOut{Initialize}{Initialize}
\SetKwInOut{Procedure}{Procedure}
\Input{ $ \mathcal{T}, \mathcal{R},\mathcal{G}, m_{a,b,a',b'}, CR, P_{SEN}, P_{RX} $ \;}
\Initialize{ $e_{0,a,b}, l_{r,0,a,b}, u_{0,t}, b_{r,0}, P_{TX,a,b} = f(a,b), P_{move_{a,b,a',b'}} = f(m_{a,b,a',b'}) $\; }
\Procedure{}
      SOLVE \textit{OROS} ($ \mathcal{T}$) \;
      GET $e_{t,a,b}, l_{r,t,a,b}, u_{r,t}, b_{r,t} \forall t \in \mathcal{T}$\;
      \Output{ $e_{t,a,b}, l_{r,t,a,b}, u_{r,t}, b_{r,t} , \qquad \forall t \in \mathcal{T}$ \;}
\caption{Offline Oracle}
\label{alg:offline}
\end{algorithm}

\begin{algorithm}[!t]
\small
\SetKwInOut{Input}{Input}
\SetKwInOut{Output}{Output}
\SetKwInOut{Return}{return}
\SetKwInOut{Initialize}{Initialize}
\SetKwInOut{Procedure}{Procedure}
\SetKwInOut{Goto}{Go To}
\Input{$ \mathcal{T}, \mathcal{R},\mathcal{G},  CR, P_{SEN}, P_{RX}, \mathcal{W}, W $ \;}
\Initialize{ $e_{0,a,b}, l_{r,0,a,b}, u_{0,t}, b_{r,0}, m_{a,b,a',b'} \leftarrow \emptyset, P_{move_{a,b,a',b'}} = f(m_{a,b,a',b'}), t=0, P_{TX,a,b} = f(a,b), \mathcal{W}\subseteq  \mathcal{T} $ \; }
\Procedure{}
      \While{ $t < |\mathcal{T}|$}{
      UPDATE $\mathcal{W} \leftarrow  \mathcal{W} = \{t, \dots, t+W\}$  \;
        SOLVE \texttt{OROS:} OROS($\mathcal{W}, e_{t,a,b}, l_{r,t,a,b}$ ) \;
        GET $e_{t,a,b}, l_{r,t,a,b}, u_{r,t}, b_{r,t} \qquad \forall t \in \mathcal{W}$ \;
         \For{$i \in \mathcal{W}$}{
             MOVE Robot according to $l_{r,i,a,b}$ \;
             UPDATE $m_{a,b,a',b'}$ \;
             UPDATE $P_{move_{a,b,a',b'}} \leftarrow $ Eq.~\ref{eq:Pmove} \;
             UPDATE $P_{TX,a,b} \leftarrow $ Eq.~\ref{eq:PTX}\;
             $t = t + 1 $\;
             \If{Obstacle}{
              UPDATE $\mathcal{W} \leftarrow  \mathcal{W} = \{t, \dots, t+W\}$  \;
              \textbf{Go To} \texttt{OROS} \;
             } 
         } 
      }
      \Output{ $ e_{t,a,b}, l_{r,t,a,b}, u_{r,t}, b_{r,t} \qquad \forall t \in \mathcal{T}$ \;}
\caption{Online OROS}
\label{alg:online}
\end{algorithm}

\section{Performance Evaluation}
\label{sec:perf_eval}

In this section, we evaluate the performance of the above \name{} optimization module when dealing with collaborative robot scenarios. We first introduce the main robot and architectural software components characterizing our scenario. Then, in order to evaluate the orchestration benefit, we compare the performances of the online \name{} approach  against the offline \benchmark{}. To this aim, robot battery and window size $W$ must be carefully evaluated and configured. As the overall exploration performances are strictly related with the obstacles' distribution, we generate a series of random scenarios to provide general insights.

\subsection{Robot Overview}
\begin{figure}
    \centering
    \begin{subfigure}[t]{0.5\columnwidth}
         \centering
         \includegraphics[width=\textwidth]{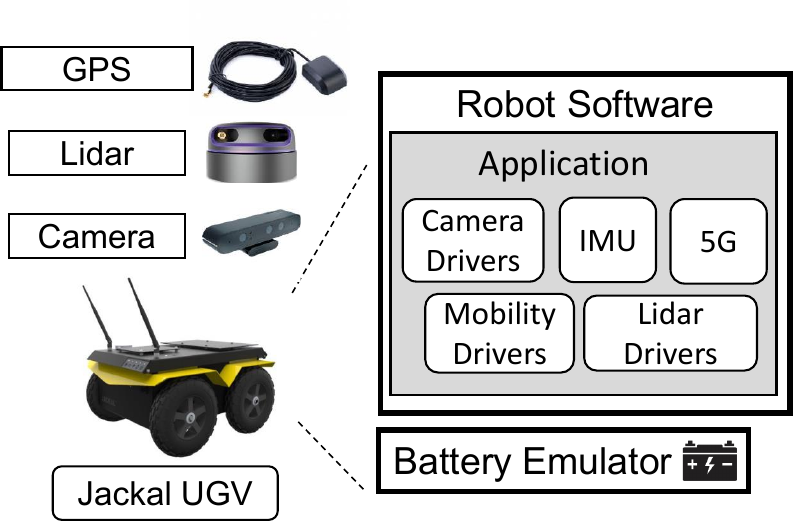}
     \end{subfigure}
         \begin{subfigure}[t]{0.45\columnwidth}
         \centering
         \includegraphics[width=\textwidth]{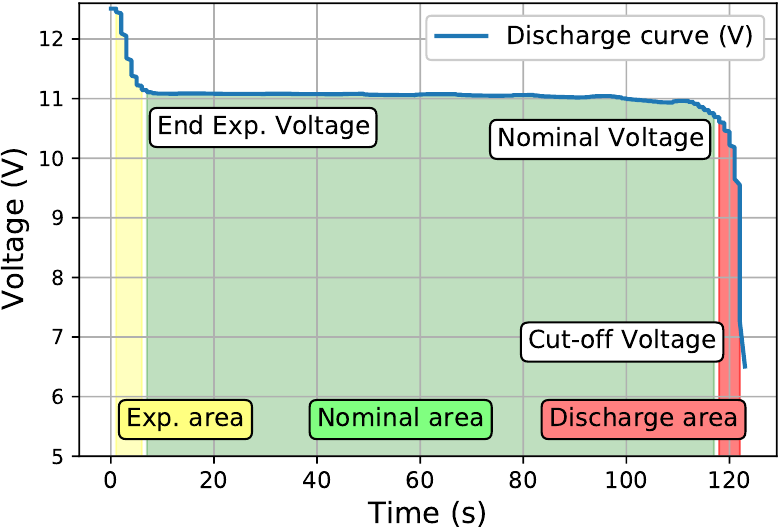}
     \end{subfigure}
    \caption{Robot software components and non-linear battery discharge example.}
    \label{fig:robot_architecture}
\end{figure}

We consider a set of Jackal UGV terrestrial mobile robots\footnote{Additional technical specifications are available at: https://clearpathrobotics.com/jackal-small-unmanned-ground-vehicle/} for outdoor applications in hard-to-reach environments, and rely on publicly available open-source code implementing the main software components\footnote{Available at: https://github.com/jackal}.
The virtual robot instances are provided with a camera, a Global Positioning System (GPS) device, an Inertial Measurement Unit (IMU), a LiDAR sensor and 5G peripherals.
Fig.~\ref{fig:robot_architecture} presents the main robot simulated components adopted along our experiments.
The ROS software composing the robot functionalities is composed by multiple modules interacting to each other by means of topic message exchange. The Gazebo software exploits these messages to simulate the robot's movement and the sensors' behavior in the virtual environment.
The Robot State Machine (RSM) incorporates the computation of estimated current consumption of each component, and manages the switching of robot sensors and communication, as well as the robot navigation and speed according to the \name{} commands.
In particular, upon receiving a new exploration target, the RSM ROS node calls the \emph{move\_base} ROS service which, in turn, calculates a global and local navigation path plans for the robot motion.
By default, ROS networks parameters such as the navigation maximum speed and lidar activation are static. In the context of our work, as discussed in Sec.~\ref{sec:problem}, \name{} can determine the use of sensors and the robot navigation. Therefore, we make use of the \emph{dynamic\_reconfigure} ROS service within our experiments to enable the RSM ROS node to modify the parameters accordingly.
During the exploration phase, the robot continuously transmits data through the 5G network. The update messages include detected obstacles by the lidar, battery status, and camera video stream. Upon reaching an exploration goal, the robot waits for new instructions. The process repeats until reaching the full discovery of the target area.

In order to validate the performance of \name{} in realistic scenarios, we considered a non-linear robot battery discharge model and implement it within our experiments, integrating a dedicated software module within the Gazebo simulator\footnote{Available at: https://github.com/nilseuropa/gazebo\_ros\_battery}.
Conversely to standard discharge models which simply consider the charge dropping in a linear fashion with respect to the current drawn from the virtual battery and regardless of the environment temperature, our implementation considers the exponential charge behavior typical of chemical batteries at different temperatures.
The module requires the three main additional components: a battery joint link, the definition of the energy consumers and a configuration file in \emph{yaml} format. 
We add the battery joint link to the Unified Robot Description Format (URDF) file which includes the mathematical model describing the battery behavior in our gazebo robot model, and include the new energy consumers in the RSM. Then, we set the main characteristics of our battery in the configuration file. An example of a battery discharge curve resulting from our experiments is depicted in Fig.~\ref{fig:robot_architecture}, considering a $20^{\circ}$C temperature and a nominal voltage of $10.62$V.
We consider four main sources of energy consumption: motion, data transmission, data reception and sensing. A ROS topic with $1$ second periodicity is used to track the instantaneous current consumption of each component, which, together with the actual battery voltage, is used to estimate the power consumption of each element. Similarly, an additional ROS topic is used to account for the effects of the battery recharge.

\begin{figure*}
    \centering
    \includegraphics[clip, trim = 0cm 0cm 0cm 0cm, width=1.5\columnwidth]{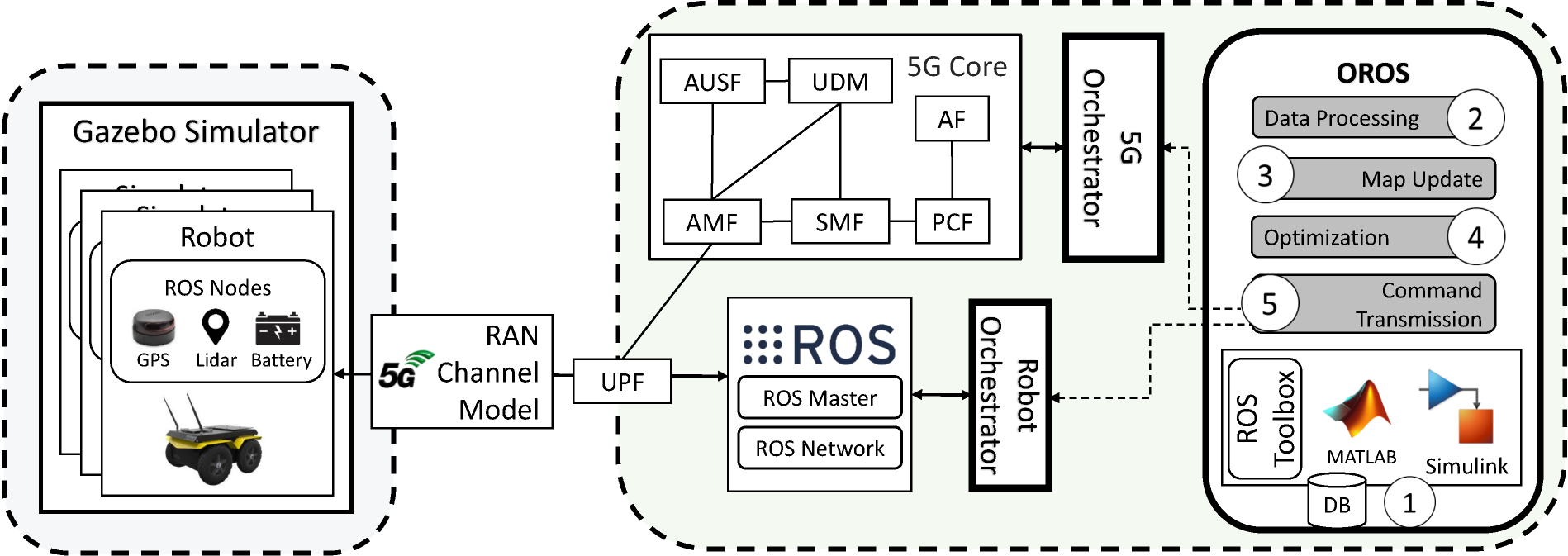}
    \caption{Testbed Architecture}
    \label{fig:system_architecture}
\end{figure*}

\subsection{Testbed Framework and \name{} Workflow}
\label{subsec:system_archi}

The following subsection describes the software and hardware architecture implementing \name{}.
We adopted two different machines, one acting as edge platform running Matlab/Simulink software and co-located with the \name{} real-time orchestrator, and another running as robot client, hosting the ROS middleware and the Gazebo robot simulator.
The two machines are connected by means of a simulated 5G channel link~\cite{TR38.900}, allowing IP-based communication among the main operational entities.
Combined with the usage of 5G, Optimization and ROS toolboxes, Simulink acts as interface for the \name{}, translating the outcome of the online/offline optimization into topic messages, and communicating them through the ROS Master and the existing ROS network finally reaching the exploration robots.
The ROS Master acts as main coordinator of the topic exchange, enabling communication over the IP-based ROS network for topics related to sensing, motion, and behavior, which are interpreted by the Gazebo simulator for realistic emulation and visual representation. The communication paradigm follows the ROS default publish/subscribe method.
Fig.~\ref{fig:system_architecture} illustrates the overall system architecture and main software components adopted along the experiments. 

More in details, the \name{} workflow can be decomposed into the following main steps:

\textbf{1. Data Reception}
Along their exploration task, robots periodically report to \name{} the information about their status within an unknown environment. This data consists of the instantaneous position and orientation of the robot, as well as the State of Charge (SOC) of the batteries and camera/LiDAR data at the reporting time instant. Input data provided by the robots pass through the simulated 5G link towards a series of ROS topic subscribers listening by \name{}.

\textbf{2. Data Processing}
 In order to optimize the robot path and sensor planning, \name{} evaluates the incoming robot messages to identify the achievement of exploration goals, the presence of obstacles, or the need for new exploration goals. In scenarios where robots do not need to transmit data, the orchestrator does not consider them for data evaluation. Additionally, when one or more robots run out of battery power, \name{} passes over such robots in subsequent time intervals when performing path planning operation.

\textbf{3. Map Update}
\name{} maintains a global exploration map by joining the information collected by the swarm of robots.
If a new obstacle has been detected, a new path planning will be enforced even if that obstacle does not interfere with the actual path plan. Such a policy is encouraged by the fact that each obstacle significantly affects the reachability of a neighboring grid element. 

\textbf{4. Optimization}
It is of key importance to keep the exploration goals updated with the latest obstacle information as to avoid unaware robots to collide with objects, or waste energy into unnecessary exploration steps. Therefore, when an update message of a robot device includes information regarding unknown obstacles, a new path plan is computed by solving an online instance of the problem described in Sec.~\ref{sec:problem}, accounting for the latest values reported by the robots.

\textbf{5. Command Generation and Transmission}
Upon defining a new path plan, the set of instructions are converted into exploration goal commands and delivered to the robots by means of the simulated 5G link. 

Finally, if the area has been completely explored, or the simulation reaches the maximum number of epochs $|\mathcal{T}|$ allowed for the area exploration purposes, the simulation ends.

\subsection{Scenario Setup and Methodology}
\label{sec:scenarioSetup}
In our work, we consider a variable set of robots that need to be jointly orchestrated as to explore an unknown area in the shortest time. Table~\ref{tab:setup} summarizes the experimental parameters used in our evaluation. 

All robots start their exploration from the same point within the area of interest, co-located with the charging station. 
We set $\Delta t = 10$s and $\Delta x = 10 m$, which translates into $|g_{a,b}| = 100m^2$. This allows to derive the velocity of the robot as $\frac{\Delta_x}{\Delta_t} = 1 m/s$ if it moves into the up/down/left/right direction, or at 
$\frac{\sqrt{2}\Delta_x}{\Delta_t} = \sqrt{2} m/s$ when moving in diagonal, which is in line with the maximum speed of $2 m/s$ provided by commercial robot devices~\cite{Robot_velocity}.
Each robot is equipped with a fully charged battery at the beginning of the exploration phase.
About the charging station, in our experiment we consider the worst case scenario with a single charging station providing the lowest charging rate among the possible commercial options, i.e., $CR=9.24 J/s$, as in the case of TurtleBot3\footnote{https://www.robotis.us/lipo-battery-charger-lbc-010/}. 
\begin{table}[t]
\caption{Experimental Setup}
\label{tab:setup}
\centering
\begin{tabular}{ll|ll}
\hline
\textbf{Definition}	& \textbf{Value} & \textbf{Definition}	& \textbf{Value}	\\
\hline
$|\mathcal{T}|$		& 16 & $|\mathcal{R}|$		& 1,2,3\\
$A \times B$                        & $40\times40$ $m^2$ & $|g_{a,b}|$ & $10\times10$ $m^2$ \\
$B_{max}$ & $4500$ J & $\Delta t$	        & $10$ s \\
$CR$	        & $9.24$ J/s & $v$	        & $1$, $\sqrt{2}$ m/s \\
$P_{SEN}$	    & $12$ W  &   $P_{RX}$	    & $4$ W  \\
$NF_{RX}$	            & $4$ dB & $N_{0}$       & $-174$ dBm/Hz \\
$G_{A_{Rx}}$  & $10$ dB   & $G_{A_{Tx}}$   & $10$ dB\\ 
\hline
\end{tabular}
\vspace{-4mm}
\end{table} 
We characterize the energy consumed by the robot for locomotion, sensor operation and radio wireless communication as follows.
First, the power consumed by the locomotion $P_{move_{a,b,a',b'}}$ (see Equation~\ref{eq:Pmove}) depends on the terrain-velocity constant $m_{a,b,a',b'}$, which also takes into account detected obstacles. This constant can be defined as: 
\begin{equation}
  m_{a,b,a',b'} =
    \begin{cases}
      0         & \text{if $a' = a$ and $b' = b$}\\
      \infty    & \text{if there is an obstacle}\\
                & \text{between $(a,b)$ and $(a',b')$ }\\
      v = \{1,\sqrt{2}\}   & \text{if $a' \neq a$ or $b' \neq b$}
    \end{cases}       
\end{equation}
where $v$ is the velocity in m/s.
Note that we assume no power consumption for turning the robot in the desired direction.
Second, we characterize the power consumed by the sensors $P_{SEN}$, which includes processing all the generated data by the camera, LiDAR and other sensors, as well as the power consumed by the local computing infrastructure. Based on the values from~\cite{Mei2005}, we set it to take a value of $12W$. 

Moreover, we consider each robot is equipped with a 5G New Radio (NR) antenna module, consuming energy whenever transmitting or receiving data. 
The power dissipated by the robot depends on its distance from the serving base station and on the corresponding Signal to Noise Ratio (SNR) which, in turn, affects the adoption of a particular MCS.
The processing power consumed to encode radio packets is also affected by the instantaneous MCS in use. A detailed characterization of the power consumption in the virtual base station as a function of the MCS can be obtained, e.g., from~\cite{Ayala2021}.
We leverage the work of~\cite{5Gmodel} to model realistically the energy consumption for uplink (UL) transmission at the robot side, for every transmission time interval (TTI). It considers the power consumption from the Radio Frequency (RF) chain and the baseband (BB) processing, focusing on RF transmitted power, Signal-to-Noise-Ratio (SNR), path loss, transmission gains and losses, and real-time traffic demand.
In particular, the path loss can be derived as~\cite{ITUR2019}:
\begin{gather}
	PathLoss(a,b) = 20log_{10}(d(a,b))+20log_{10}(f)-147.55,
\end{gather}
where $d(a,b)$ is the distance between the robot and the BS (expressed in meters), and $f$ is the carrier frequency, which we set to $3.5$ GHz within our experiments~\cite{TR38.901}.
The noise is considered for the transmission performance, which depends on the receiver noise ($NF_{RX}$), the thermal noise ($N_0$) and the number of RBs ($N_{RBS}$) used to accommodate the traffic demand in a single TTI:
\begin{gather}
	Noise = N_0 + NF_{RX} + 10·log_{10}(12\cdot2^\mu\cdot15kHz \cdot N_{RBS}),
\end{gather}
where $12$ is the number of subcarriers in every resource block considering normal cyclic prefix, $\mu$ is the 5G numerology index which goes from $0$ to $6$ as specified in~\cite{38.211}, and $N_{RBS}$ is dynamically derived depending on the channel statistics and instantaneous traffic demand, following the specifications detailed in 3GPP TS 38.214~\cite{TS38.214}. We assume $\mu=0$. The transmitted power of the RF chain $P_{Tx RF}$ is calculated as:
\begin{gather}
\label{eq:PTX}	
 P_{Tx,a,b} = SNR_{Target} - G_{A_{Tx}} + L_{C_{Tx}} - G_{A_{Rx}} + \\ L_{C_{Rx}} + PathLoss(a,b) + Noise, \notag
\end{gather}
where ${G_{A_{Tx}}}$ and ${G_{A_{Rx}}}$ are the antenna gains at transmitter and receiver, i.e., robot and BS, respectively, ${L_{C_{Tx}}}$ and ${L_{C_{Rx}}}$ are the corresponding cable losses, and $SNR_{Target}$ is the target SNR at the receiver side.
Based on the results and findings of~\cite{5GUplink}, we assume the transmission power consumption of the BB processing $P_{Tx BB}$ to be approximately 2.12 W, and set ${L_{C_{Tx}}}$ and ${L_{C_{Rx}}}$ $0$dB and $0$dB, respectively, and assign ${G_{A_{Tx}}}$ and ${G_{A_{Rx}}}$ $10$dB and $10$dB, respectively.
In case of downlink transmission (DL), we consider an average value of $4W$ according to the measurements of~\cite{Ayala2021}, and assume the receiving module always on, to enable reachability of the robot devices with control messages at any time.
We use the Gurobi solver~\cite{gurobi} to address the optimization problem described in Sec.~\ref{sec:problem}.

\subsection{Scenario Characterization and Sensitivity Analysis}
\label{sec:Sensitivity}

In our case study, there are some key parameters that have a significant impact on the robot exploration performances. For example, the achievable exploration rate is primarily affected by the robot battery capacity and by the total size of the target area, and secondarily by the deployed number of robots, which allow exploring the same area in a collaborative manner.
Perhaps surprisingly, also the number of obstacles in the area as well as their localization alter the exploration performances, and finally, the window size parameter has a great impact on the orchestrator optimization and computing time, which affects the delay between two consecutive navigation instructions.
In fact, smaller window sizes might provide faster solutions, but may also lead to situations where the robot can not travel to unexplored area units within the given time frame, finally prioritizing the battery savings and preferring a standing position. We define such occurrences as \emph{corner cases}.
For all these reasons, in the following we perform a sensitivity analysis on several key-parameters of our model formulation, assessing the impact of their variation on the overall system performances.

\begin{figure}[t]
    \centering
    \includegraphics[clip, trim = 0cm 0cm 0cm 0cm, width=\columnwidth]{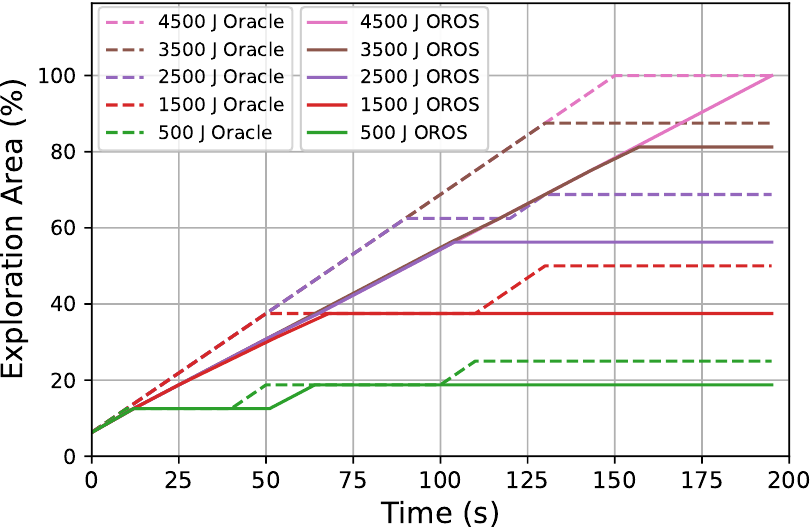}
    \caption{Exploration performances for different battery sizes and orchestration approaches.}
     \vspace{-3mm}
    \label{fig:battery_selection}
\end{figure}

\begin{figure*}
     \centering
     \begin{subfigure}[t]{0.32\textwidth}
         \centering
         \includegraphics[width=\textwidth]{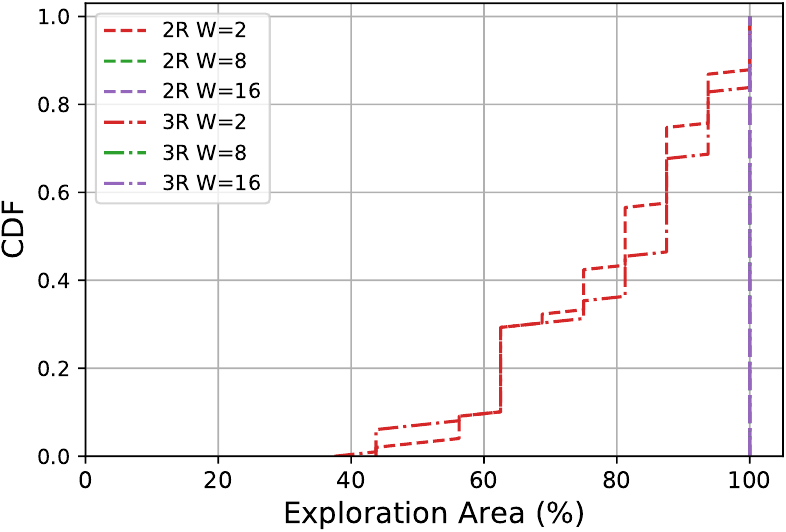}
         \caption{$3$ Obstacles}
         \vspace{3mm}
         \label{fig:a}
     \end{subfigure}
     \begin{subfigure}[t]{0.32\textwidth}
         \centering
         \includegraphics[width=\textwidth]{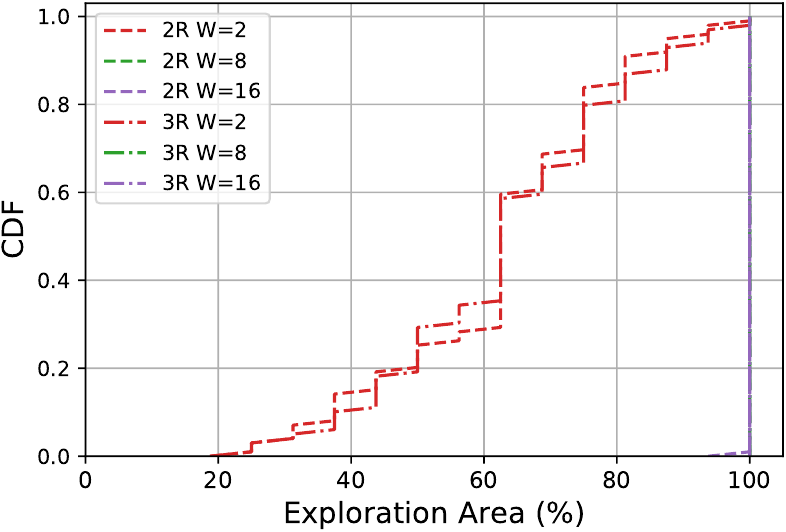} 
         \caption{$5$ Obstacles}
         \label{fig:b}
     \end{subfigure}
     \begin{subfigure}[t]{0.32\textwidth}
         \centering
         \includegraphics[width=\textwidth]{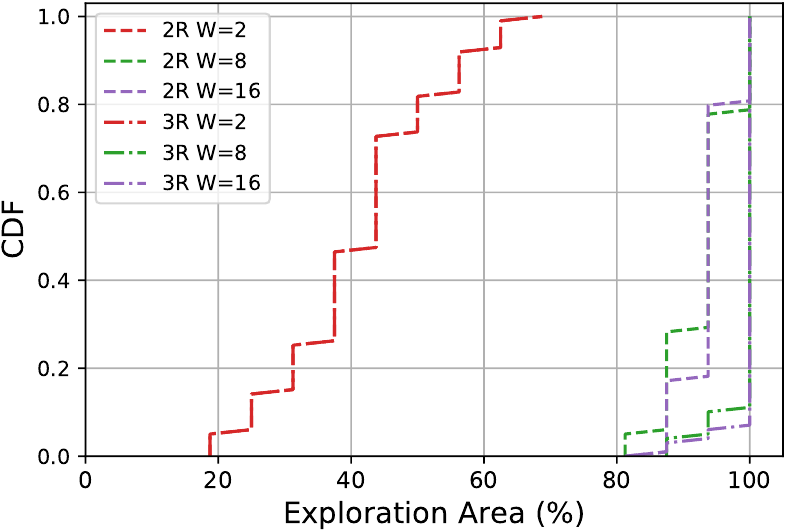}
         \caption{$9$ Obstacles}
         \label{fig:c}
     \end{subfigure}
     \vspace{3mm}
     
    \begin{subfigure}[t]{0.32\textwidth}
         \centering
         \includegraphics[width=\textwidth]{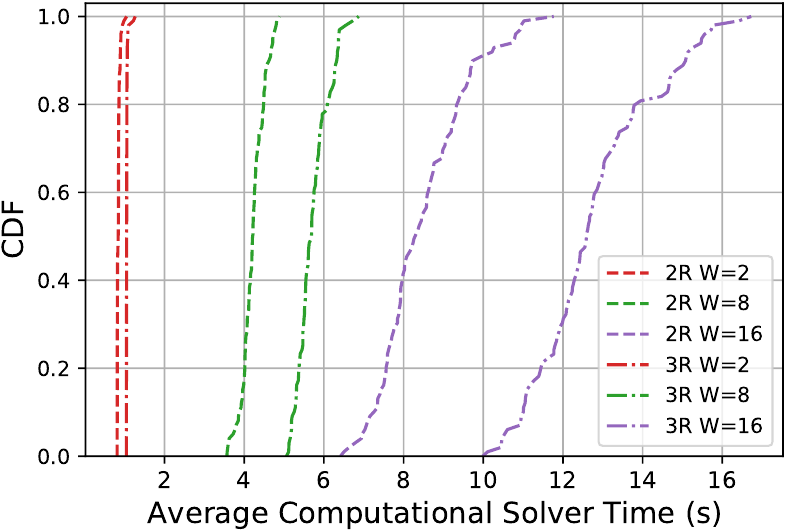}
         \caption{$3$ Obstacles}
         \label{fig:d}
     \end{subfigure}
     \begin{subfigure}[t]{0.32\textwidth}
         \centering
         \includegraphics[width=\textwidth]{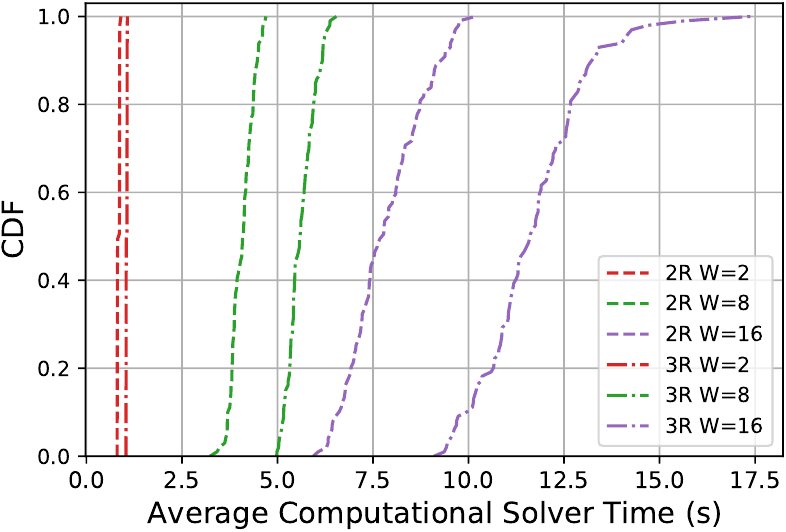}
         \caption{$5$ Obstacles}
         \label{fig:e}
     \end{subfigure}
     \begin{subfigure}[t]{0.32\textwidth}
         \centering
         \includegraphics[width=\textwidth]{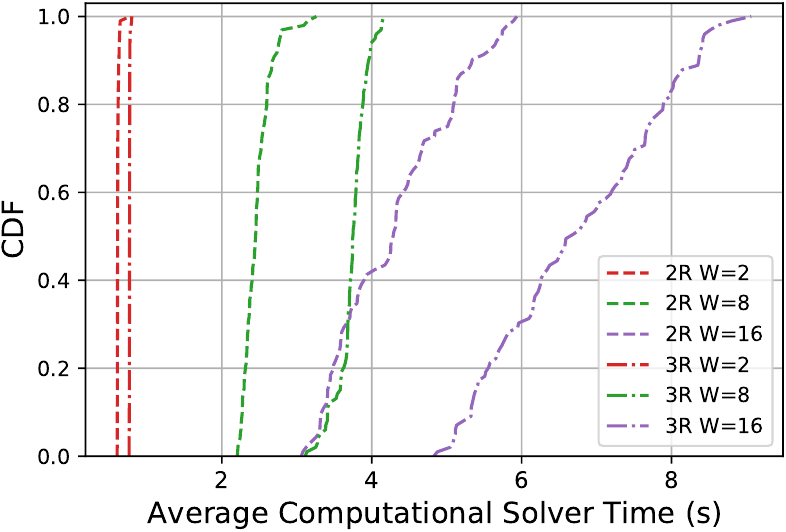}
         \caption{$9$ Obstacles}
         \label{fig:f}
     \end{subfigure}
     \vspace{3mm}
     \caption{Exploration performances and average solver time for different number of obstacles and variable window sizes.}
     \label{fig:window_exploration}
     \vspace{-3mm}
\end{figure*}

\textbf{Battery Size} Fig.~\ref{fig:battery_selection} depicts the explored area (in percentage) over time when one robot with different battery sizes ($B_{max}$) tries to explore an unknown area. For each battery size, we provide two lines, one for the orchestrator expected offline Oracle results (dashed line) and another for the actual online \name{} results (solid lines) when using the Gazebo simulator.
In both cases, the highest battery capacity (4500J) has a steady incremental behavior and can reach the totality of the explored area at the end of the simulation. On the contrary, lower battery capacities, such as 500J, 1000J and 1500J, achieve the worst performances at the end of the simulation. The curves raise while the robot explores new area units, while remains steady for a period of time when the robot is sent to recharge its battery.
The two curves present very different behaviors over time. 
In particular, we can notice a clear performance gap (and delay) for the \name{} approach.
This behavior is mainly due to the more realistic and precise assumptions taken by the \name{} approach, which also considers acceleration, deceleration, stop and turn as causes of energy consumption.
In the following, if not else specified, we select a battery capacity of $2500$J, which allows exploring the majority of the target area while exploiting the possibility of recharge.

\textbf{Window Size} 
In Fig.~\ref{fig:window_exploration} we collect the results of $100$ simulations, testing a variable number of obstacles randomly placed in our reference scenario, as well as different decision window sizes $W$ ranging from $2$ to $16$ time steps. We consider up to three robots within this evaluation. The positioning of the obstacles was carried out in such a way as to avoid the repetition of the scenario and guarantee access to the entire exploration area. The upper row depicts the exploration results at the end of the simulations, expressed as the percentage of the total explored area. As expected, deploying more robots increases the achievable exploration rate independently of the optimization window size used. Nevertheless, when considering the same number of robot, larger $W$ values positively affect the overall performances, with $W=8$ and $W=16$ achieving similar performances due to the relatively small target area. 
Conversely, the lower row evaluates the solver execution time required to compute an optimal solution in each decision step. It should be noted that the number of calls to the solver depends on the window size and the number of obstacles identified by the robots in the reference scenario. Therefore, we  consider the computation time averaged over the number of calls to the solver.
From the plots, it can be observed that the solver computing time depends on both the number of robots and window size $W$. This clearly impacts on the time between the reception of two consecutive navigation goals.
Interestingly, the computing time decreases along the exploration task, as the occurrence of new obstacles brings additional constraints to the optimization model, finally breaking possible symmetries and easing the overall solving process.
Considering the exploration performances as well as the comparison between computational solving times for the different window sizes, within our settings, a window size of $8$ brings the best trade-off between computational solving times and exploration performances, and it is therefore used in the following of the paper.

\begin{figure}[t]
    \centering
    \includegraphics[clip, trim = 0cm 0cm 0cm 0cm, width=0.8\columnwidth]{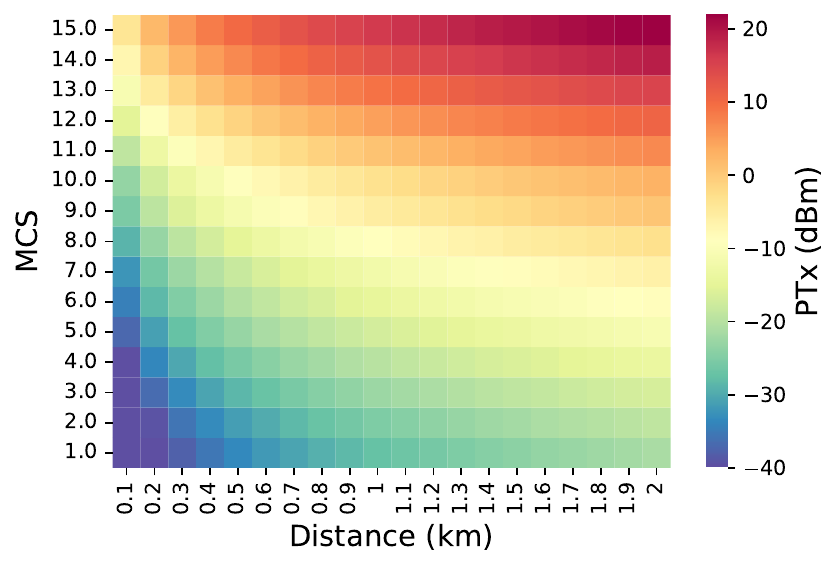}
    \caption{Transmission energy consumption for different MCS values and communication range.}
    \label{fig:snr_energy2}
    \vspace{-3mm}
\end{figure}

\setcounter{figure}{10}
\begin{figure*}[t!]
     \begin{subfigure}{0.33\textwidth}
         \centering
         \includegraphics[width=\columnwidth]{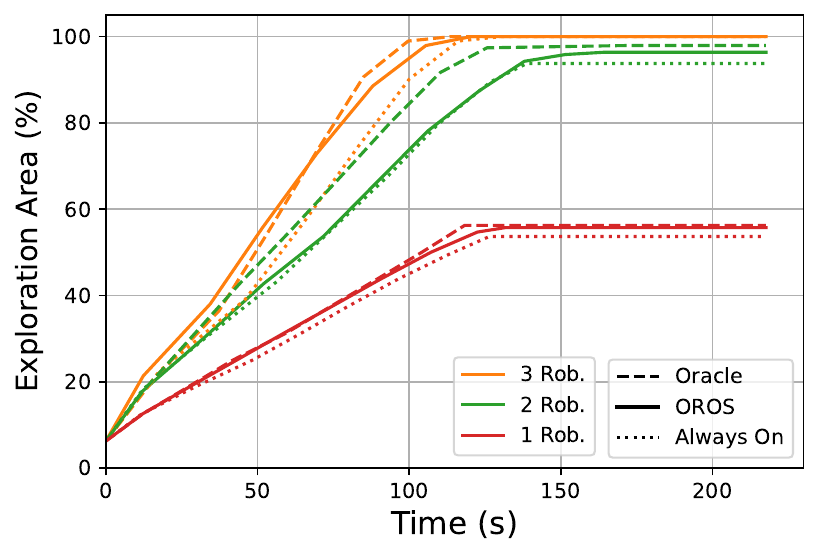}
         \caption{$3$ Obstacles}
         \label{fig:11a}
     \end{subfigure}
     \begin{subfigure}{0.33\textwidth}
         \centering
         \includegraphics[width=\columnwidth]{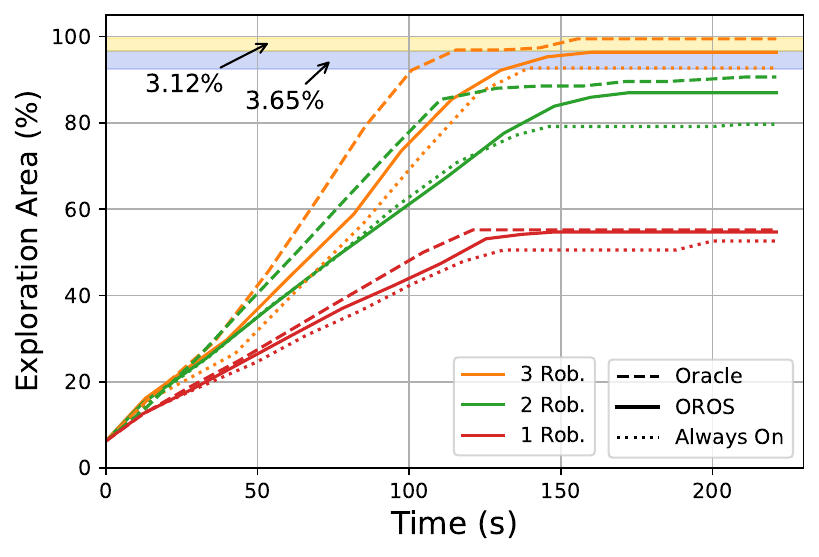}
         \caption{$5$ Obstacles}
    \label{fig:exploration_results_5}
    \end{subfigure}
     \begin{subfigure}{0.33\textwidth}
         \centering
         \includegraphics[width=\columnwidth]{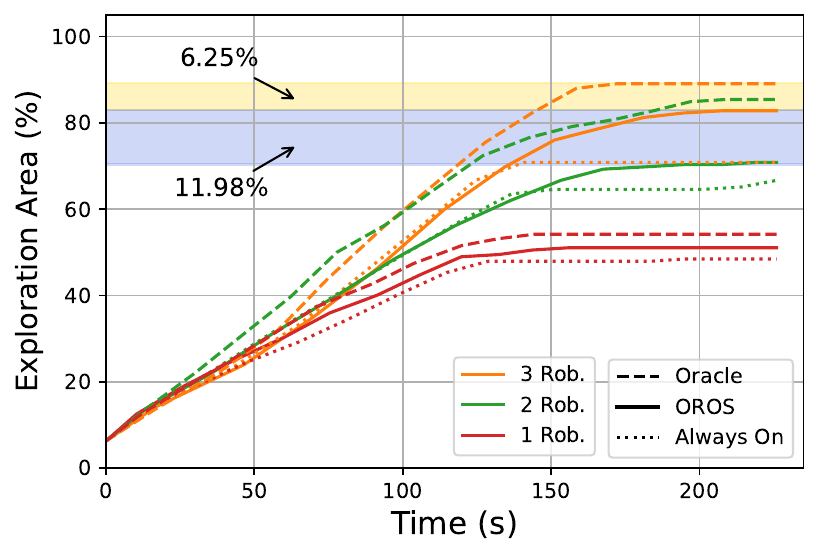}
         \caption{$9$ Obstacles}
         \label{fig:exploration_results_9}
     \end{subfigure}
     \vspace{3mm}
     \caption{Exploration area over time using Online OROS, Offline Oracle and Always On approaches}
     \label{fig:exploration_results}
     \vspace{-3mm}
\end{figure*}

\setcounter{figure}{9}
\begin{figure}[ht]
    \centering
    \includegraphics[clip, trim = 0cm 0cm 0cm 0cm, width=0.95\columnwidth]{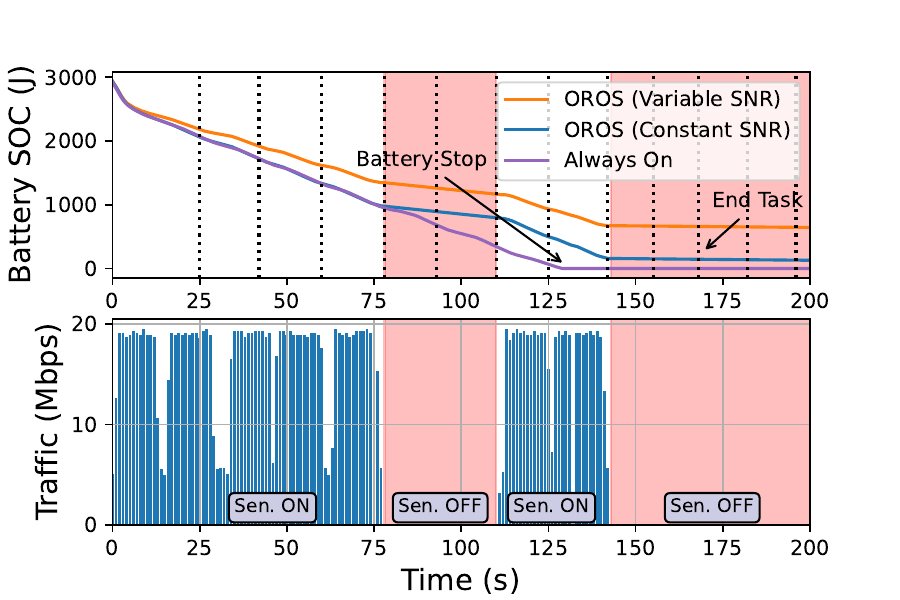}
    \caption{Impact of OROS on the robot energy consumption.}
    \vspace{-3mm}
    \label{fig:final_consumption_comparison}
\end{figure}

\textbf{Channel Conditions} Variable 5G channel conditions affect the energy consumption of the robot, as more intensive modulation and demodulation processing is required to keep the communication active. To characterize this effect on the battery consumption, in Fig.~\ref{fig:snr_energy2} we consider the traffic volume generated by one robot towards the edge platform and derive the energy consumption related to transmission over the 5G interface following the model described in Sec.~\ref{sec:scenarioSetup}. In particular, we consider the impact of path-loss on the required transmission power as described in Eq.~\ref{eq:PTX}, following the work of~\cite{5GUplink}. From this test study, we notice how the heatmap gradient reflects an increase in power transmission required to achieve larger SNR values and, in turn, the possibility to adopt higher Modulation and Coding Schemes (MCS), as function of the distance from the robot and the serving base station.

\subsection{Assessing the gain of the \name{} orchestrator}

In the scenario considered in our study, the 5G and robot orchestration cooperate to devise the best robot energy-saving strategy while performing exploration tasks.

Fig.~\ref{fig:final_consumption_comparison} characterizes the impact of smart orchestration decisions in a multi-robot deployment, focusing on a single robot and its battery SOC over time, as well as its uplink traffic. 
In the figure, we compare \name{} against a baseline approach that considers an \emph{Always On} scenario, and the approach considered in~\cite{Delgado22}, dubbed as \emph{OROS Constant SNR}, which considers the transmission energy consumption with a constant channel quality level.
Within these settings, the set of robot navigation goals is kept constant to ease the comparison, while the red area highlights the time instants where \name{} imposes the decision of switching off the robot's sensors. In the same plot, dashed black lines identify the occurrence of a new command from \name{} within the experiment.

We remark that as shown in Sec.~\ref{sec:scenarioSetup}, variable wireless channel conditions impose different load on the local processing unit due to lighter (or heavier) modulation\slash demodulation tasks and larger bandwidth utilization. 
The battery SOC depicted in the upper plot presents an exponential behavior at the beginning of the experiment, which comes from the adoption of a realistic non-linear battery model implemented in our Gazebo simulator. Furthermore, we can notice, within the highlighted time segments, how smart orchestration decisions allow energy saving in the robot battery. In particular, between $75$ and $110$ seconds, the robot is required to go through an already explored area. \name{} promptly reacts and informs the robot to keep its sensors off, pursuing energy optimization.
The energy that \name{} allows to save enables the robot to continue the exploration of new areas, in contrast with the \emph{always On} case, which consumes all its power and stops before task completion.
Finally, Fig.~\ref{fig:exploration_results} considers the scenario introduced in Sec.~\ref{sec:Sensitivity} comparing the performances of \emph{Online OROS} (solid lines) against two baseline approaches such as \emph{Oracle} (dashed lines), and \emph{Always On} (dotted lines), implemented within the Gazebo simulator.
We consider up to three robots equipped with a $2500$J battery, a variable set of obstacles, namely $3$, $5$ and $9$, a window size of $8$ time steps (W=8), and a maximum simulation time limit of $200$ seconds. Each line is averaged over $12$ experiments, each one characterized by a random obstacle distribution.
From the picture, it can be noticed how the exploration rate increases faster by using a larger number of robots for a fixed  number of obstacles. 
At the same time, as previously discussed, a larger number of obstacles may translate into multiple update messages from the robots. As each update message containing one or several undiscovered obstacles triggers a new optimization task, this increases the overall computation time as well as the time robots have to wait for obtaining the next navigation goal. When comparing \emph{OROS} with the state-of-the-art \emph{Always On} approach, we can notice how the orchestration of the robot and network applications results in energy savings that allows extending the exploration range, with improvements up to $11.98$\% within our considered scenario.
In general, the larger the number of deployed robots, the more likely a unit area can be crossed by multiple robots, thus enabling larger energy savings when employing our proposed \name{} strategy. Similarly, increasing the number of obstacles in a given scenario limits the navigation options, thus forcing the robots to follow similar paths and revisit already explored areas, e.g., in presence of corridors or dead-ends, favoring larger energy savings. 
For this reason, given the a-priori knowledge of obstacle location available in the \emph{Offline Oracle} approach, the latter is used as benchmark for exploration rate and navigation in each simulation instance.
When comparing \emph{Online OROS} with the optimal \emph{Offline Oracle}, we can notice a $3.12$\% and $6.25$\% exploration rate gap in the case of $5$ and $9$ obstacles, respectively, for the $3$ robot scenarios. 
The \emph{Offline Oracle} avoids any replanning that may occur in the online version due to e.g., dead-ends and corners. 
Additionally, conversely from \emph{Offline Oracle} that pre-computes the navigation path of robots before execution (such time is not considered in Fig.~\ref{fig:exploration_results}), \emph{Online OROS} requires constant update messages from the robots. This translates into multiple optimization solver calls to derive and update the best path planning, slowing down the initial exploration phase. 
Once again, obstacle density on the map has a direct impact on the \emph{Online OROS} and \emph{Offline Oracle} performances, that in fact present minimal to no differences in the $3$ obstacles scenario as depicted in Fig.~\ref{fig:11a}.

\subsection{Discussion}
In this work, we presented a framework for the deployment of outdoor robotic applications exploiting a 5G infrastructure. We implemented and evaluated an optimization approach combining the orchestration of the 5G mobile infrastructure with the energy-aware optimization of the on-board robot sensors and drivers. 
First, we performed a sensitivity analysis of the different model parameters. Later, we assessed the potential gain of the OROS orchestrator. 
To do so, we considered a realistic non-linear discharge rate for the battery, aware of the fact  that its behavior in real scenarios will be affected by multiple factors, including the type of battery used, operational temperatures, and battery wear.
Moreover, an exhaustive evaluation of the benefits of the orchestration solution should consider the overhead brought by virtualization software on the robot platform, both in terms of communication overhead and computational burden.
Additionally, the current MILP formulation suffers scalability issues when increasing the number of robots, limiting its applicability in small fleet scenarios. Such an aspect may be mitigated by adopting heuristic algorithms and pre-solving schemes that consider contextual knowledge.
Nevertheless, starting from our test scenario and implementation, we believe our findings can help characterize the expected behavior of complex systems.

\section{Conclusions and Future Work}
\label{sec:conclusion}

Due to limited computing and energy resource availability, cloud-based robot deployments rely on mobile infrastructure to enable collective robotic intelligence exchange and increase their collaboration and efficiency when performing tasks. However, current solutions are limited by the fact that robot operating systems and ICT computing and communication platforms, do not have means to interact with each other.
In this paper, we propose a joint optimization framework for the concurrent orchestration of the robotic, computing and communication infrastructure domains. Our results show that collaborative real-time robot operations would benefit from the adoption of \name{}, a joint orchestration solution that significantly improves energy consumption and task completion duration of 5G-enabled robots.
Future works include the implementation of the proposed \name{} framework in a real deployment, comprising off-the-shelf outdoor robots and an operational 5G mobile infrastructure. 
In order to fully characterize the benefits of our orchestration solution, we will also explore more realistic non-flat environments and large-scale deployments involving multiple edge platforms, which may bring different battery depletion times and further exacerbate the problem-solving.

\section*{Acknowledgment}
The research leading to these results has been supported by the Spanish Ministry of Economic Affairs and Digital Transformation and the European Union – NextGeneration EU, in the framework of the Recovery Plan, Transformation and Resilience (PRTR) (Call UNICO I+D 5G 2021, ref. number TSI-063000-2021-6), by the CERCA Programme from the Generalitat de Catalunya, and by the European Union's H$2020$ 5G ERA Project (grant no. $101016681$).

\bibliographystyle{IEEEtran}
\bibliography{biblio}

\begin{IEEEbiography}[{\includegraphics[width=1in,height=1.25in,clip,keepaspectratio]{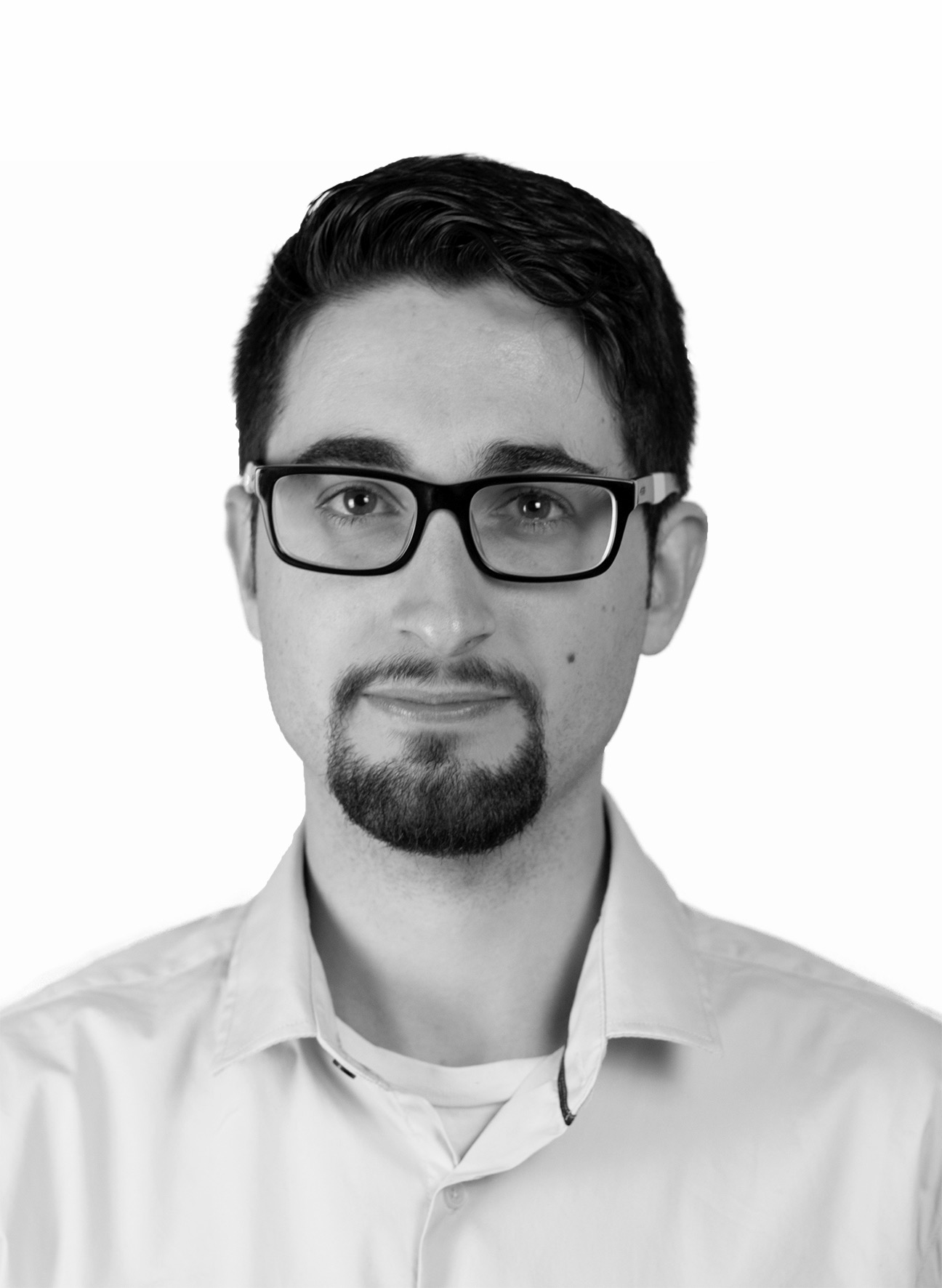}}]{Arnau Romero} received his M.Sc. in Industrial Engineering specialisation in Automatic Control in 2022 from the Universitat Politècnica de Catalunya, Spain. He is currently enrolled as PhD candidate in Automatics, Robotics and Computer Vision at the Universitat Politècnica de Catalunya in Barcelona. He works as junior researcher at i2CAT Foundation, Barcelona, Spain. His research interests include robotics, energy modeling and efficiency, machine learning, computer vision and their applicability to 5G.
\end{IEEEbiography}

\vspace{-15mm}

\begin{IEEEbiography}[{\includegraphics[width=1in,height=1.25in,clip,keepaspectratio]{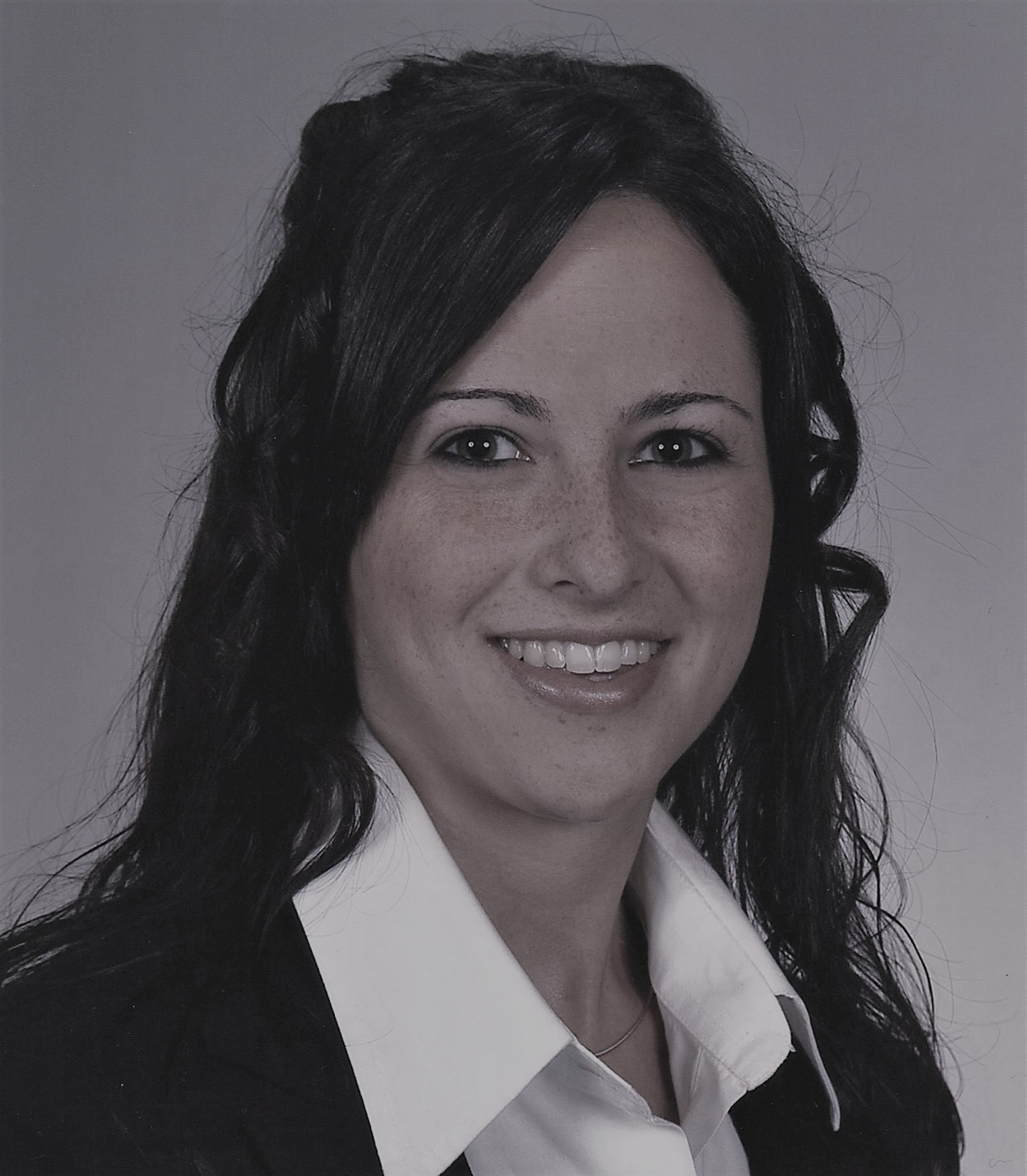}}]{Carmen Delgado} received the MSc degree in telecommunications engineering, the M.Sc. degree in biomedical engineering, and the PhD degree (cum laude) in mobile network information and communication technologies from the University of Zaragoza, Spain, in 2013, 2014, and 2018, respectively. In 2018, she was a postdoctoral researcher with the Internet Technology and Data Science Lab, University of Antwerp, associated with IMEC, Belgium. She  works as senior researcher at i2CAT Foundation. Her main research interests lie in the field of wireless sensor networks, Internet of Things, mobile networks, resource allocation, battery-less sensors and communications and Artificial Intelligence of Things.

\end{IEEEbiography}

\vspace{-15mm}
\begin{IEEEbiography}[{\includegraphics[width=1in,height=1.25in,clip,keepaspectratio]{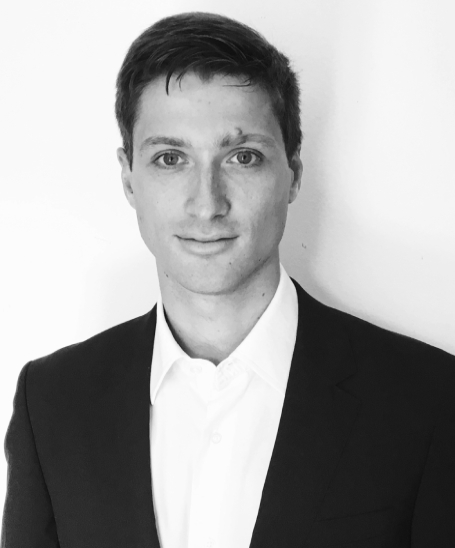}}]{Lanfranco Zanzi}(S'17--M'22) received his B.Sc. and M.Sc. in Telecommunication Engineering from Polytechnic University of Milan (Italy) in 2014 and 2017, respectively, and the Ph.D. degree from the Technical University of Kaiserslautern (Germany) in 2022. He works as senior research scientist at NEC Laboratories Europe. His research interests include network virtualization, machine learning, blockchain, and their applicability to 5G and 6G mobile networks in the context of network slicing.
\end{IEEEbiography}

\vspace{-15mm}

\begin{IEEEbiography}[{\includegraphics[width=1in,height=1.25in,clip,keepaspectratio]{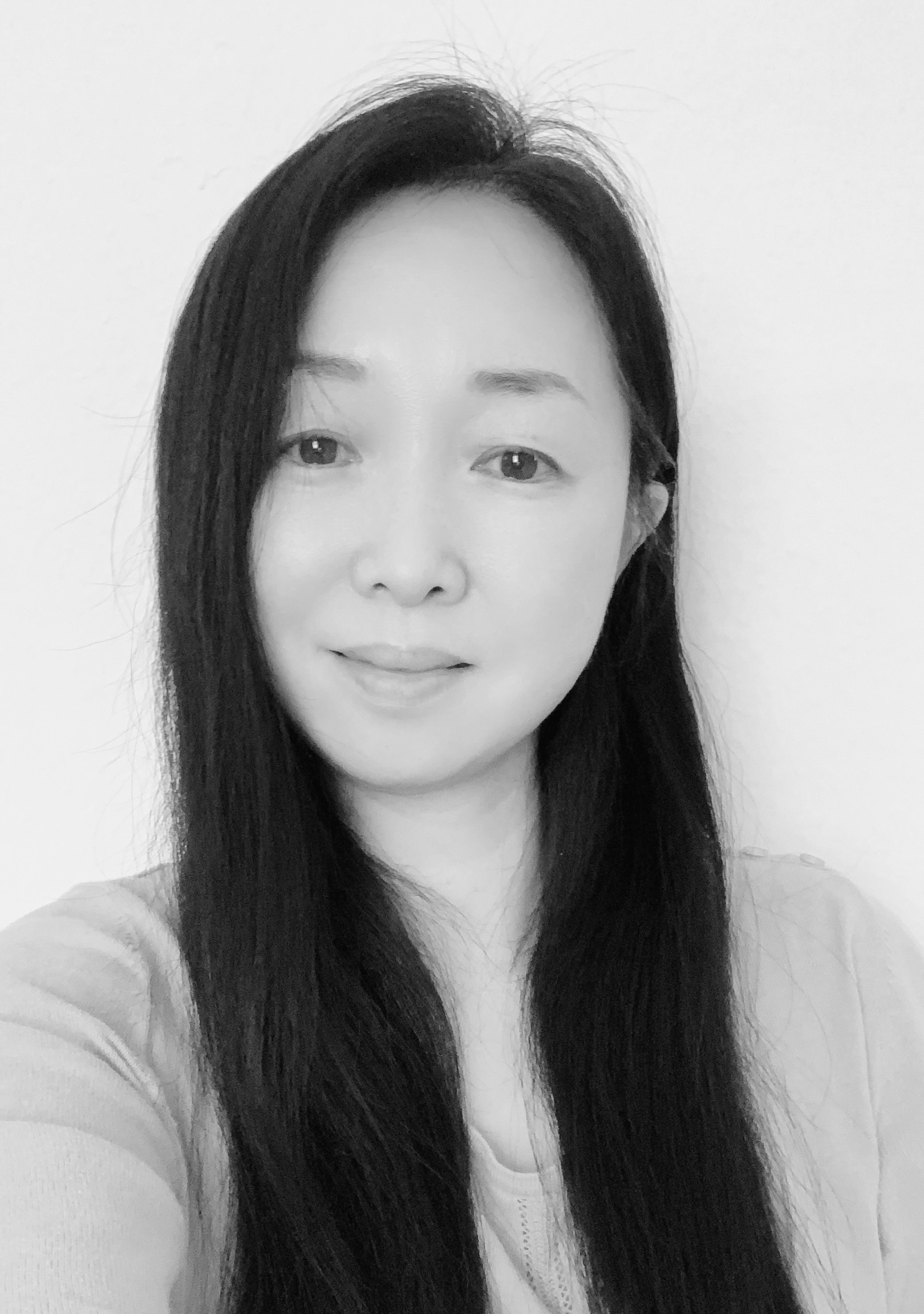}}]{Xi Li} is a Senior Researcher on 6G Networks R\&D at NEC Laboratories Europe, Germany, and the Vice Chairman of the 5GPPP Architecture Working Group. She received her M.Sc. in 2002 from the Technical University of Dresden and Ph.D. in 2009 from University of Bremen, Germany. She was the Technical Led of EU H2020 5Growth project and the work package led of EU H2020 5G-Crosshaul and 5G-TRANSFORMER projects. Previously, she was a senior researcher fellow and lecturer at the University of Bremen and a solution designer at Telefonica, Germany. She has published 80+ journal and conference publications, and given many invited talks in various industrial events and international conferences. She is an inventor of 18 patents including 7 granted ones, and active in contributing to IETF CCAMP WG with two published RFCs and received best overall award at IETF’99 Hackathon in 2017. Her research interests comprise the design for next generation mobile and wireless networks, open and virtualized RAN, distributed edge platform solutions, applying AI/ML for resource and service management and automation. 
\end{IEEEbiography}

\vspace{-15mm}

\begin{IEEEbiography}[{\includegraphics[width=1in,height=1.25in,clip,keepaspectratio]{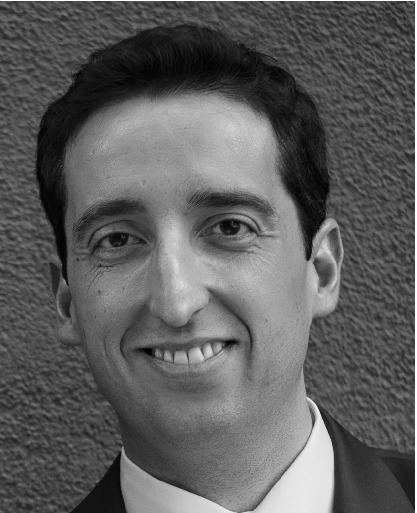}}]{Xavier Costa-P\'erez} (M'06--SM'18) is Head of Beyond 5G Networks R\&D at NEC Laboratories Europe, Scientific Director at the i2Cat R\&D Center and Research Professor at ICREA. His team contributes to products roadmap evolution as well as to European Commission R\&D collaborative projects and received several awards for successful technology transfers. In addition, the team contributes to related standardization bodies: 3GPP, ETSI NFV, ETSI MEC and IETF. Xavier has been a 5GPPP Technology Board member, served on the Program Committee of several conferences (including IEEE Greencom, WCNC, and INFOCOM), published at top research venues and holds several patents. He also serves as Editor of IEEE Transactions on Mobile Computing and Transactions on Communications journals. He received both his M.Sc. and Ph.D. degrees in Telecommunications from the Polytechnic University of Catalonia (UPC) in Barcelona and was the recipient of a national award for his Ph.D. thesis.
\end{IEEEbiography}

\end{document}